\documentclass{article}

\usepackage{amsmath}
\usepackage{amssymb}
\usepackage{amsthm}
\usepackage{enumerate}
\usepackage{graphicx,color}
\usepackage{pst-tree}
\usepackage{subfigure}
\usepackage{url}
\usepackage{xspace}

\newcommand{\nc}{\newcommand}

\nc{\logspace}{{\sc logspace}\xspace}
\nc{\nlogspace}{{\sc nlogspace}\xspace}
\nc{\ptime}{{\sc ptime}\xspace}
\nc{\np}{{\sc np}\xspace}
\nc{\conp} {\textrm{co}{\sc np}\xspace}
\nc{\pspace}{{\sc pspace}\xspace}
\nc{\exptime}{{\sc exptime}\xspace}
\nc{\nexptime}{{\sc nexptime}\xspace}

\nc{\blogspace}{{\scriptsize {\bf L}}}
\nc{\bnlogspace}{{\scriptsize {\bf  NL}}}
\nc{\bptime}{{\scriptsize {\bf PTIME}}}
\nc{\bnp}{{\scriptsize {\bf NP}}}
\nc{\bconp}{{\scriptsize {\bf coNP}}}
\nc{\bpspace}{{\scriptsize {\bf PSPACE}}}
\nc{\bexptime}{{\scriptsize {\bf EXPTIME}}}
\nc{\bnexptime}{{\scriptsize {\bf NEXP}}}

\nc{\cH}{{\mathcal H}}
\nc{\cM}{{\mathcal M}}
\nc{\cO}{{\mathcal O}}
\nc{\cS}{{\mathcal S}}
\nc{\cT}{{\mathcal T}}
\nc{\cC}{{\mathcal C}}

\nc{\Dom}{\text{Dom}}
\nc{\XPath}{\text{XPath}}

\nc{\enc}{\text{enc}}
\nc{\integers}{\mathbb{Z}}
\nc{\lab}{\text{lab}}
\nc{\nat}{\mathbb{N}}
\nc{\ot}{\leftarrow}
\nc{\out}{\text{out}}
\nc{\rat}{\mathbb{Q}}
\nc{\reals}{\mathbb{R}}
\nc{\rhs}{\text{rhs}}
\nc{\depth}{\text{depth}}

\nc{\desc}{//}
\nc{\<}{\langle}

\newtheorem{theorem}{Theorem}[section]
\newtheorem{corollary}[theorem]{Corollary}
\newtheorem{proposition}[theorem]{Proposition}
\newtheorem{lemma}[theorem]{Lemma}

\newtheorem{defn}[theorem]{Definition}

\newtheorem{exmp}[theorem]{Example}

\newenvironment{example}{\begin{exmp} \rm}{\end{exmp}}
\newenvironment{definition}{\begin{defn} \rm}{\end{defn}}

\begin{document}

    \title{On Typechecking Top-Down XML Tranformations: Fixed Input or
      Output Schemas\thanks{An extended abstract of a part of this paper
      appeared as Section~3 in reference~\cite{martensnevenpods04} in
      the ACM SIGACT-SIGMOD-SIGART Symposium on Principles of Database
      Systems, 2004.}}

  \author{Wim Martens\thanks{Corresponding author. Email:
      wim.martens@uhasselt.be} \and Frank Neven\thanks{Email:
      frank.neven@uhasselt.be} \and Marc Gyssens\thanks{Email:
      marc.gyssens@uhasselt.be}}

    \date{ \small
      Hasselt University and \\
      Transnational University of Limburg,\\
      Agoralaan, Gebouw D\\
      B-3590 Diepenbeek, Belgium
    }

    \maketitle
    
    \begin{abstract} 
      Typechecking consists of statically verifying whether the output of
      an XML transformation always conforms to an output type for
      documents satisfying a given input type.  
      In this general setting,
      both the input and output schema as well as the transformation are
      part of the input for the problem.  However, scenarios where the
      input or output schema can be considered to be fixed, are quite
      common in practice. In the present work, we investigate the
      computational complexity of the typechecking problem in the latter
      setting.
    \end{abstract}

\section{Introduction}

In a typical XML data exchange scenario on the web, a user community
creates a common schema and agrees on producing only XML data
conforming to that schema.  This raises the issue of typechecking:
verifying at compile time that every XML document which is the result
of a specified query or document transformation applied to a valid
input document satisfies the output
schema~\cite{suciudbpl01,suciusigmodrecordtc}.

The typechecking problem is determined by three parameters: the
classes of allowed input and output schemas, and the class of
XML-transformations.  As typechecking quickly becomes
intractable~\cite{alon2journal,martensneventcs05,msvjournal}, we focus
on simple but practical XML transformations where only little
restructuring is needed, such as, for instance, in filtering of
documents.  In this connection, we think, for example, of
transformations that can be expressed by structural
recursion~\cite{suciuvldbunql} or by a top-down fragment of
XSLT~\cite{bmnjournal}.  As is customary, we abstract such
transformations by unranked tree
transducers~\cite{manneven,martensneventcs05}.  As schemas, we adopt
the usual Document Type Definitions (DTDs) and their robust
extensions: regular tree languages~\cite{msvjournal,leeXML} or,
equivalently, specialized
DTDs~\cite{vianupapakpods00,vianupapakicdt03full}. The latter serve as a
formal model for XML Schema~\cite{w3schema}.

Our work studies sound and complete typechecking algorithms, an
approach that should be contrasted with the work on general-purpose
XML programming languages like XDuce~\cite{xduce} and
CDuce~\cite{cduce}, for instance, where the main objective is fast and
sound typechecking. The latter kind of typechecking is always
incomplete due to the Turing-completeness of the considered
XML-transformations. That is, it can happen that type safe
transformations are rejected by the typechecker.  As we only consider
very simple transformations which are by no means Turing-complete, it
makes sense to ask for complete algorithms.

The typechecking scenario outlined above is very general: both the
schemas and the transducer are determined to be part of the input.
However, for some exchange scenarios, it makes sense to consider the
input and/or output schema to be fixed when transformations are always
from within and/or to a specific community.  Therefore, we revisit the
various instances of the typechecking problem considered
in~\cite{martensneventcs05} and determine the complexity in the
presence of fixed input and/or output schemas.  The main goal of this
paper is to investigate to which extent the complexity of the
typechecking problem is lowered in scenarios where the input and/or
output schema is fixed.  An overview of our results is presented in
Table~\ref{tab:tc-complexities}.

The remainder of the paper is organized as follows. In
Section~\ref{sec:relatedwork}, we discuss related work. In
Section~\ref{sec:definitions}, we provide the necessary definitions.
In Section~\ref{sec:fixed}, we discuss typechecking in the restricted
settings of fixed output and/or input schemas.  The results are
summarized in Table~\ref{tab:tc-complexities}. 
We obtain several new cases for which typechecking is in polynomial
time: \emph{(i)} when the input schema is fixed and the schemas are
DTDs with SL-expressions; \emph{(ii)} when the output schema is fixed
and the schemas are DTDs with NFAs; and \emph{(iii)} when both the
input and output schemas are fixed and the schemas are DTDs using
DFAs, NFAs, or SL-expressions. We conclude in
Section~\ref{sec:conclusion}.

\section{Related Work} \label{sec:relatedwork}

The research on typechecking XML transformations was initiated by
Milo, Suciu, and Vianu~\cite{msvjournal}. They obtained the
decidability for typechecking of transformations realized by
$k$-pebble transducers via a reduction to satisfiability of monadic
second-order logic.  Unfortunately, in this general setting, the
latter non-elementary algorithm cannot be improved~\cite{msvjournal}.
Interestingly, typechecking of $k$-pebble transducers has recently
been related to typechecking of compositions of macro tree
transducers~\cite{engmanpeb}. Alon et
al.~\cite{alon1journal,alon2journal} investigated typechecking in the
presence of data values and show that the problem quickly turns
undecidable. As our interest lies in formalisms with a more manageable
complexity for the typechecking problem, we choose to work with XML
transformations that are much less expressive than $k$-pebble
transducers and that do not change or use data values in the process
of transformation.

A problem related to typechecking is type
inference~\cite{ms_pods99,vianupapakpods00}. This problem consists in
constructing a tight output schema, given an input schema and a
transformation. Of course, solving the type inference problem implies
a solution for the typechecking problem, namely, checking containment
of the inferred schema into the given one. However, characterizing
output languages of transformations is quite
hard~\cite{vianupapakpods00}. For this reason, we adopt different
techniques for obtaining complexity upper bounds for the typechecking
problem.

The transducers considered in the present paper are restricted
versions of the DTL-programs, studied by Maneth and
Neven~\cite{manneven}.  They already obtained a non-elementary upper
bound on the complexity of typechecking (due to the use of monadic
second-order logic in the definition of the transducers).  Recently,
Maneth et \mbox{al.}  considered the typechecking problem for an
extension of DTL-programs and obtained that typechecking was still
decidable~\cite{manseidlpods05}. Their typechecking algorithm, like
the one of \cite{msvjournal}, is based on inverse type inference. That
is, they compute the pre-image of all ill-formed output documents and
test whether the intersection of the pre-image and the input schema is
empty.  Tozawa considered typechecking with respect to tree automata
for a fragment of top-down XSLT~\cite{tozawa}. He uses a more general
framework, but he was not able to derive a bound better than
double-exponential on the complexity of his algorithm.

Martens and Neven investigated polynomial time fragments of the
typechecking problem by putting syntactical restrictions on the tree
transducers, and making them as general as
possible~\cite{martensnevenjcss05}.  Here, tractability of the
typechecking problem is obtained by bounding the \emph{deletion path
  width} of the tree transducers. The deletion path width is a notion
that measures the number of times that a tree transducer copies part
of its input. In particular, it also gives rise to tractable fragments
of the typechecking problem where the transducer is allowed to delete
in a limited manner.

\section{Preliminaries} \label{sec:definitions}

In this section we provide the necessary background on trees,
automata, and tree transducers. In the following, $\Sigma$ always
denotes a finite alphabet. 

By $\nat$ we denote the set of natural numbers. A \emph{string} $w =
a_1\cdots a_n$ is a finite sequence of $\Sigma$-symbols. The set of
positions, or the \emph{domain}, of $w$ is $\Dom(w)=\{1,\ldots,n\}$.
The length of $w$, denoted by $|w|$, is the number of symbols
occurring in it. The label $a_i$ of position $i$ in $w$ is denoted by
$\lab^w(i)$.  The size of a set $S$, is denoted by $|S|$.

As usual, a {\em nondeterministic finite automaton} (NFA) over
$\Sigma$ is a tuple $N = (Q,\Sigma,\delta,I,F)$ where $Q$ is a finite
set of states, $\delta:Q\times\Sigma\to 2^Q$ is the transition
function, $I\subseteq Q$ is the set of initial states, and $F\subseteq
Q$ is the set of final states. A {\em run} $\rho$ of $N$ on a string
$w\in\Sigma^*$ is a mapping from $\Dom(w)$ to $Q$ such that
$\rho(1)\in \delta(q,\lab^w(1))$ for $q\in I$, and for
$i=1,\ldots,|w|-1$, $\rho(i+1)\in\delta(\rho(i),\lab^w(i+1))$.  A run
is \emph{accepting} if $\rho(|w|)\in F$. A string is \emph{accepted}
if there is an accepting run. The language accepted by $N$ is denoted
by $L(N)$.  The \emph{size} of $N$ is defined as $|Q|+|\Sigma|+
\sum_{q\in Q,a\in\Sigma}|\delta(q,a)|$.

A \emph{deterministic finite automaton} (DFA) is an NFA where
\emph{(i)} $I$ is a singleton and \emph{(ii)} $|\delta(q,a)|\leq 1$
for all $q\in Q$ and $a\in\Sigma$.

\subsection{Trees and Hedges}
It is common to view XML documents as finite trees with labels from a
finite alphabet $\Sigma$. 
Figures~\ref{fig:xml-a} and~\ref{fig:xml-b} give an example of an XML
document together with its tree representation.  Of course, elements
in XML documents can also contain references to nodes. But, as XML
schema languages usually do not constrain these nor the data values at
leaves, it is safe to view schemas as simply defining tree languages
over a finite alphabet. In the rest of this section, we introduce the
necessary background concerning XML schema languages.

\newbox\subfigbox
\makeatletter
\newenvironment{subfloat}
{\def\caption##1{\gdef\subcapsave{\relax##1}}
\let\subcapsave=\@empty
\let\sf@oldlabel=\label
\def\label##1{\xdef\sublabsave{\noexpand\label{##1}}}
\let\sublabsave\relax
\setbox\subfigbox\hbox
\bgroup}
{\egroup
\let\label=\sf@oldlabel
\subfigure[\subcapsave]{\box\subfigbox}}
\makeatother

\begin{figure}[!h]
  \centering
  \begin{subfloat}
    \begin{minipage}{10cm}
\begin{verbatim}
<store>
  <dvd>
    <title> "Amelie" </title>
    <price> 17 </price>
  </dvd>
  <dvd>
    <title> "Good bye, Lenin!" </title>
    <price> 20 </price>
  </dvd>
  <dvd>
    <title> "Pulp Fiction" </title>
    <price> 11 </price>
    <discount> 6 </discount>
  </dvd>
</store>
\end{verbatim}
    \end{minipage}
    \caption{An example XML document.\label{fig:xml-a}}
  \end{subfloat}
  \subfigure[Its tree representation with data values.\label{fig:xml-b}]{
    \pstree[nodesep=2pt,levelsep=0.6cm,treesep=0.3cm]{\TR{store}}{
      \pstree{\TR{dvd}}{
        \pstree{\TR{title}}{
          \TR{``Amelie''}
        }
        \pstree{\TR{price}}{
          \TR{17}
        }
      }
      \pstree{\TR{dvd}}{
        \pstree{\TR{title}}{
          \TR{``Good bye, Lenin!''}
        }
        \pstree{\TR{price}}{
          \TR{20}
        }
      }
      \pstree{\TR{dvd}}{
        \pstree{\TR{title}}{
          \TR{``Pulp Fiction''}
        }
        \pstree{\TR{price}}{
          \TR{11}
        }
        \pstree{\TR{discount}}{
          \TR{6}
        }
      }
    }
  }
\caption{An example of an XML document and its
  tree representation.\label{fig:xml}}
\end{figure}

\begin{figure}[h]
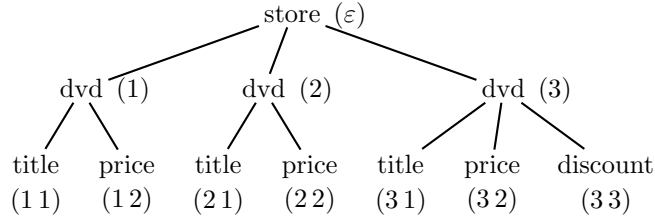
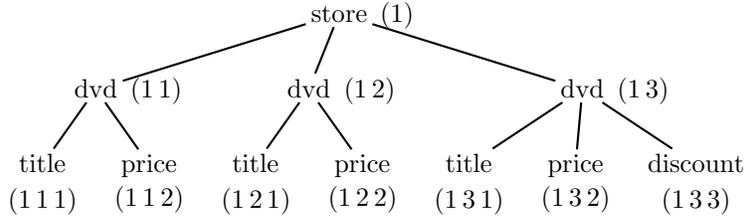

  \centering
  \subfigure[The tree of Figure~\ref{fig:xml-b} without data values. The nodes are
  annotated next to the labels, between brackets.\label{fig:xml-c}]{
    \pstree[nodesep=2pt,levelsep=1cm,treesep=0.5cm]{\TR{store}~[tnpos=r]{($\varepsilon)$}}{
      \pstree{\TR{dvd}~[tnpos=r]{$(1)$}}{
        \TR{title}~[tnpos=b]{$(1\,1)$}
        \TR{price}~[tnpos=b,tnsep=2.6pt]{$(1\,2)$}
      }
      \pstree{\TR{dvd}~[tnpos=r]{$(2)$}}{
        \TR{title}~[tnpos=b]{$(2\,1)$}
        \TR{price}~[tnpos=b,tnsep=2.6pt]{$(2\,2)$}
      }
      \pstree{\TR{dvd}~[tnpos=r]{$(3)$}}{
        \TR{title}~[tnpos=b]{$(3\,1)$}
        \TR{price}~[tnpos=b,tnsep=2.6pt]{$(3\,2)$}
        \TR{discount}~[tnpos=b]{$(3\,3)$}
      }
    }
  }
  \subfigure[Tree of Figure~\ref{fig:xml-c} viewed as a hedge. The nodes are
  annotated next to the labels, between brackets.\label{fig:xml-d}]{
    \pstree[nodesep=2pt,levelsep=1cm,treesep=0.5cm]{\TR{store}~[tnpos=r]{$(1)$}}{
      \pstree{\TR{dvd}~[tnpos=r]{$(1\,1)$}}{
        \TR{title}~[tnpos=b]{$(1\,1\,1)$}
        \TR{price}~[tnpos=b,tnsep=2.6pt]{$(1\,1\,2)$}
      }
      \pstree{\TR{dvd}~[tnpos=r]{$(1\,2)$}}{
        \TR{title}~[tnpos=b]{$(1\,2\,1)$}
        \TR{price}~[tnpos=b,tnsep=2.6pt]{$(1\,2\,2)$}
      }
      \pstree{\TR{dvd}~[tnpos=r]{$(1\,3)$}}{
        \TR{title}~[tnpos=b]{$(1\,3\,1)$}
        \TR{price}~[tnpos=b,tnsep=2.6pt]{$(1\,3\,2)$}
        \TR{discount}~[tnpos=b]{$(1\,3\,3)$}
      }
    }
  }
  \caption{The document of Figure~\ref{fig:xml} without data values,
    viewed as a tree and as a hedge.\label{fig:xml2}}
\end{figure}

The set of \emph{unranked $\Sigma$-trees}, denoted by $\cT_\Sigma$, is
the smallest set of strings over $\Sigma$ and the parenthesis symbols
``('' and ``)'' such that, for $a\in\Sigma$ and $w\in
{\cT_\Sigma}^*$, $a(w)$ is in $\cT_\Sigma$. So, a tree is either
$\varepsilon$ (empty) or is of the form $a(t_1\cdots t_n)$ where
each $t_i$ is a tree. In the tree $a(t_1 \cdots t_n)$, the
subtrees $t_1,\ldots,t_n$ are attached to a root labeled $a$.  We
write $a$ rather than $a()$.  Note that there is no a priori
bound on the number of children of a node in a $\Sigma$-tree; such
trees are therefore \emph{unranked}. For every $t \in \cT_\Sigma$, the
\emph{set of tree-nodes} of $t$, denoted by $\Dom_T(t)$, is the set
defined as follows:
\begin{enumerate}[(i)]
\item if $t = \varepsilon$, then $\Dom_T(t) = \emptyset$; and,
\item if $t = a(t_1 \cdots t_n)$, where each $t_i \in
  \cT_\Sigma$, then $\Dom_T(t) = \{\varepsilon\} \cup \bigcup_{i=1}^n
  \{iu \mid u \in \Dom_T(t_i)\}$.
\end{enumerate}
Figure~\ref{fig:xml-c} contains a tree in which we annotated the nodes
between brackets. Observe that the $n$ child nodes of a node $u$ are
always $u1, \ldots, un$, from left to right. The \emph{label} of a
node $u$ in the tree $t = a(t_1 \cdots t_n)$, denoted by
$\lab_T^t(u)$, is defined as follows:
\begin{enumerate}[(i)]
\item if $u = \varepsilon$, then $\lab_T^t(u) = a$; and,
\item if $u = iu'$, then $\lab_T^t(u) = \lab_T^{t_i}(u')$.
\end{enumerate}
We define the \emph{depth} of a tree $t$, denoted by $\depth(t)$, as
follows: if $t = \varepsilon$, then $\depth(t) = 0$; and if $t =
a(t_1 \cdots t_n)$, then $\depth(t) = \max\{\depth(t_i) \mid
1 \leq i \leq n\} + 1$. In the sequel, whenever we say tree, we always
mean $\Sigma$-tree.  A \emph{tree language} is a set of trees.

A \emph{hedge} is a finite sequence of trees. Hence, the set of
hedges, denoted by $\cH_\Sigma$, equals $\cT_\Sigma^*$. For every
hedge $h\in \cH_\Sigma$, the \emph{set of hedge-nodes of $h$}, denoted
by $\Dom_H(h)$, is the subset of $\nat^*$ defined as follows:
\begin{enumerate}[(i)]
\item if $h=\varepsilon$, then $\Dom_H(h)=\emptyset$; and,
\item if $h=t_1 \cdots t_n$ and each $t_i\in\cT_\Sigma$, then
  $\Dom_H(h)=\bigcup_{i=1}^n \{iu\mid u\in \Dom_T(t_i)\}$. 
\end{enumerate}
The \emph{label} of a node $u = iu'$ in the hedge $h = t_1 \cdots
t_n$, denoted by $\lab_H^h(u)$, is defined as $\lab_H^h(u) =
\lab_T^{t_i}(u')$.  Note that the set of hedge-nodes of a hedge
consisting of one tree is different from the set of tree-nodes of this
tree. For example: if the tree in Figure~\ref{fig:xml-c} were to
represent a single-tree hedge, it would have the set of hedge-nodes
$\{1,11,12,13,111,112,121,\allowbreak 122,131,132,133\}$, as shown in
Figure~\ref{fig:xml-d}.  The \emph{depth} of the hedge $h = t_1 \cdots
t_n$, denoted by $\depth(h)$, is defined as $\max\{\depth(t_i) \mid
i=1,\ldots,n\}$.  For a hedge $h=t_1\cdots t_n$, we denote by
$\text{top}(h)$ the string obtained by concatenating the root symbols
of all $t_i$s, that is, $\lab_H^{t_1}(1) \cdots \lab_H^{t_n}(n)$.

In the sequel, we adopt the following conventions: we use $t, t_1,
t_2,\ldots$ to denote trees and $h, h_1, h_2,\ldots$ to denote hedges.
Hence, when we write $h=t_1\cdots t_n$ we tacitly assume that all
$t_i$'s are trees. We denote $\Dom_T$ and $\Dom_H$ simply by $\Dom$,
and we denote $\lab_T$ and $\lab_H$ by $\lab$ when it is understood
from the context whether we are working with trees or hedges.

\subsection{DTDs and Tree Automata} \label{sec:DTD-TA}

We use extended context-free grammars and tree automata to abstract
from DTDs and the various proposals for XML schemas.
We parameterize the definition of DTDs by a class of representations
$\cM$ of regular string languages such as, for instance, the class of
DFAs (Deterministic Finite Automata) or NFAs (Non-deterministic Finite
Automata). For $M\in \cM$, we denote by $L(M)$ the set of strings
accepted by $M$. We then abstract DTDs as follows.
\begin{definition}\label{ex:dtd}
  Let $\cM$ be a class of representations of regular string languages
  over $\Sigma$.  A {\em DTD} is a tuple $(d, s_d)$ where $d$ is a
  function that maps $\Sigma$-symbols to elements of $\cM$ and $s_d
  \in \Sigma$ is the start symbol.
\end{definition}
For convenience of notation, we denote $(d, s_d)$ by $d$ and leave the
start symbol $s_d$ implicit whenever this cannot give rise to
confusion.  A tree $t$ \emph{satisfies} $d$ if \emph{(i)}
$\lab^t(\varepsilon)=s_d$ and, \emph{(ii)} for every $u\in\Dom(t)$
with $n$ children, $\lab^t(u1)\cdots \lab^t(un)\in L(d(\lab^t(u)))$.
By $L(d)$ we denote the set of trees satisfying $d$.

Given a DTD $d$, we say that a $\Sigma$-symbol $a$ \emph{occurs in
  $d(b)$} when there exist $\Sigma$-strings $w_1$ and $w_2$ such that
$w_1aw_2 \in L(d(b))$.  We say that $a$ \emph{occurs in $d$} if $a$
occurs in $d(b)$ for some $b \in \Sigma$. 

We denote by DTD($\cM$) the class of DTDs where the regular string
languages are represented by elements of $\cM$. The \emph{size} of a
DTD is the sum of the sizes of the elements of $\cM$ used to represent
the function $d$.

\begin{example}
  The following DTD $(d,\text{store})$ is satisfied by the tree in
  Figure~\ref{fig:xml-c}: $$
\begin{array}{lll}
  d(\text{store}) & = & \text{dvd}\ \text{dvd}^*\\
  d(\text{dvd}) & = & \text{title}\ \text{price}\  (\text{discount} + \varepsilon)\\
\end{array}
$$
This DTD defines the set of trees where the root is labeled with
``store''; the children of ``store'' are all labeled with ``dvd''; and
every ``dvd''-labeled node has a ``title'', ``price'', and an optional
``discount'' child.
\end{example}

In some cases, our algorithms are easier to explain on well-behaved
DTDs as considered next.  A DTD $d$ is \emph{reduced} if, for every
symbol $a$ that occurs in $d$, there exists a tree $t \in L(d)$ and a
node $u \in \Dom(t)$ such that $\lab^t(u) = a$. Hence, for example,
the DTD $(d,a)$ where $d(a) = a$ is not reduced. Reducing a DTD(DFA)
is in \ptime, while reducing a DTD(SL) is in \conp (see the Appendix,
Corrollary~\ref{cor:reducing}).  Here, SL is a logic as defined next.

To define unordered languages, we make use of the specification
language SL inspired by~\cite{neventhomaswebdb99} and also used
in~\cite{alon1journal,alon2journal}. The syntax of this language is as
follows:
\begin{definition}
  For every $a\in\Sigma$ and natural number $i$, $a^{=i}$ and
  $a^{\geq i}$ are \emph{atomic SL-formulas}; ``true'' is also an
  atomic SL-formula. Every atomic SL-formula is an
  SL-formula and the negation, conjunction, and disjunction of
  SL-formulas are also SL-formulas.
\end{definition}
A string $w$ over $\Sigma$ satisfies an atomic formula $a^{=i}$ if it
has exactly $i$ occurrences of $a$; $w$ satisfies $a^{\geq i}$ if it
has at least $i$ occurrences of $a$.  Furthermore, ``true'' is
satisfied by every string. Satisfaction of Boolean combinations of
atomic formulas is defined in the obvious way.\footnote{The empty
  string is obtained as $\bigwedge_{a\in\Sigma} a^{=0}$ and the empty
  set as $\neg$ true.} By $w \models \phi$, we denote that $w$
satisfies the SL-formula $\phi$.

As an example, consider the SL-formula $\lnot(\text{discount}^{\geq 1}
\land \lnot \text{price}^{\geq 1})$.  This expresses the constraint
that a discount can only occur when a price occurs. The \emph{size} of
an SL-formula is the number of symbols that occur in it, that is,
$\Sigma$-symbols, logical symbols, and numbers (every $i$ in $a^{=i}$
or $a^{\geq i}$ is written in binary notation).

We recall the definition of non-deterministic tree automata from
\cite{bmw}. We refer the unfamiliar reader to \cite{nevenSR} for a
gentle introduction.
\begin{definition}\label{NBTA}
  A {\em nondeterministic tree automaton (NTA)\/} is a 4-tuple
  $B=(Q,\Sigma, \allowbreak \delta,F)$, where $Q$ is a finite set of
  states, $F\subseteq Q$ is the set of final states, and $\delta:
  Q\times\Sigma\to 2^{Q^*}$ is a function such that $\delta(q,a)$ is a
  regular string language over $Q$ for every $a\in\Sigma$ and $q\in
  Q$.
\end{definition}
For simplicity, we often denote the regular languages in $B$'s
transition function by regular expressions.

A \emph{run} of $B$ on a tree $t$ is a labeling $\lambda:\Dom(t) \to
Q$ such that, for every $v\in\Dom(t)$ with $n$ children,
$\lambda(v1)\cdots \lambda(vn) \in\delta(\lambda(v),\lab^t(v)).$ Note
that, when $v$ has no children, the criterion reduces to $\varepsilon
\in \delta(\lambda(v),\lab^t(v))$. A run is \emph{accepting} if the
root is labeled with an accepting state, that is,
$\lambda(\varepsilon) \in F$. A tree is accepted if there is an
accepting run. The set of all accepted trees is denoted by $L(B)$ and
is called a \emph{regular tree language}. 

A tree automaton is \emph{bottom-up deterministic} if, for all
$q,q'\in Q$ with $q\neq q'$ and $a\in\Sigma$,
$\delta(q,a)\cap\delta(q',a) = \emptyset$. We denote the set of
bottom-up deterministic NTAs by DTA.

\begin{example} \label{ex:TA-run}
  We give a bottom-up deterministic tree automaton
  $B= (Q, \Sigma,\allowbreak \delta, F)$ which accepts the parse trees of well-formed
  Boolean expressions that are true. Here, the alphabet $\Sigma$ is
  $\{\land,\lor,\lnot,\text{true},\text{false}\}$.  The states set $Q$
  contains the states $q_\text{true}$ and $q_\text{false}$, and the
  accepting state set $F$ is the singleton $\{q_\text{true}\}$. The
  transition function of $B$ is defined as follows:
  \begin{itemize}
  \item $\delta(q_\text{true},\text{true}) = \varepsilon$. We assign
    the state $q_\text{true}$ to leafs with label ``true''.
  \item $\delta(q_\text{false},\text{false}) = \varepsilon$. We
    assigns the state $q_\text{false}$ to leafs with label ``false''.
  \item $\delta(q_\text{true},\land) = q_\text{true}
    q_\text{true}^*$. 
  \item $\delta(q_\text{false},\land) = (q_\text{true} +
    q_\text{false})^* q_\text{false} (q_\text{true} +
    q_\text{false})^*$.
  \item $\delta(q_\text{true},\lor) = (q_\text{true} +
    q_\text{false})^* q_\text{true} (q_\text{true} +
    q_\text{false})^*$.
  \item $\delta(q_\text{false},\lor) = q_\text{false}
    q_\text{false}^*$.
  \item $\delta(q_\text{true},\lnot) = q_\text{false}$.
  \item $\delta(q_\text{false},\lnot) = q_\text{true}$.
  \end{itemize}

  Consider the tree $t$ depicted in Figure~\ref{fig:tree-ex-a}. The
  unique accepting run $r$ of $B$ on $t$ can be graphically
  represented as shown in Figure~\ref{fig:tree-ex-b}.  Formally, the
  run of $B$ on $t$ is the function $\lambda : \Dom(t) \to Q : u
  \mapsto \lab^r(u)$.  Note that $B$ is a DTA.

  \begin{figure}
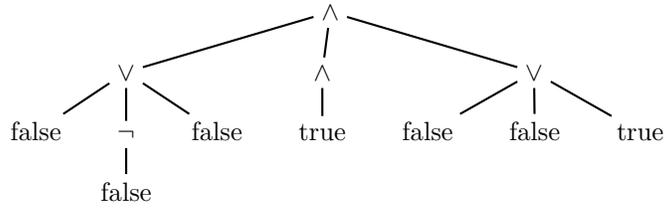
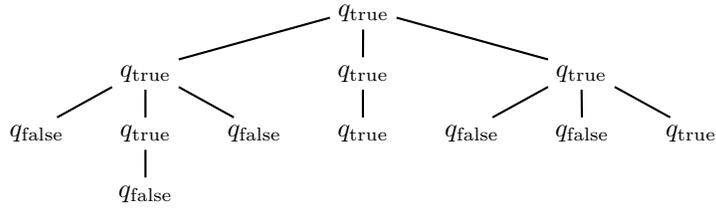

    \centering
    \subfigure[The tree $t$.\label{fig:tree-ex-a}]{
        \pstree[levelsep=0.8cm, nodesep=3pt]{\TR{$\land$}}{
          \pstree{\TR{$\lor$}}{
            \TR{false}
            \pstree{\TR{$\lnot$}}{
              \TR{false}
            }
            \TR{false}
          }
          \pstree{\TR{$\land$}}{
            \TR{true}
          }
          \pstree{\TR{$\lor$}}{
            \TR{false}
            \TR{false}
            \TR{true}
          }
        }
    }
    \qquad
    \subfigure[Graphical representation of the run $r$ of $B$ on $t$.\label{fig:tree-ex-b}]{
        \pstree[levelsep=0.8cm, nodesep=3pt]{\TR{$q_\text{true}$}}{
          \pstree{\TR{$q_\text{true}$}}{
            \TR{$q_\text{false}$}
            \pstree{\TR{$q_\text{true}$}}{
              \TR{$q_\text{false}$}
            }
            \TR{$q_\text{false}$}
          }
          \pstree{\TR{$q_\text{true}$}}{
            \TR{$q_\text{true}$}
          }
          \pstree{\TR{$q_\text{true}$}}{
            \TR{$q_\text{false}$}
            \TR{$q_\text{false}$}
            \TR{$q_\text{true}$}
          }
        }
    }
    \caption{Illustrations for Example~\ref{ex:TA-run}.}
  \end{figure}
\end{example}

As for DTDs, we parameterize NTAs by the formalism used to represent
the regular languages in the transition functions $\delta(q,a)$. So,
for a class of representations of regular languages $\cM$, we denote
by NTA$({\cM})$ the class of NTAs where all transition functions are
represented by elements of $\cM$. The \emph{size} of an automaton $B$
then is $|Q|+|\Sigma|+\sum_{q \in Q, a\in \Sigma}|\delta(q,a)|$.
Here, by $|\delta(q,a)|$, we denote the size of the automaton
accepting $\delta(q,a)$. Unless explicitly specified otherwise,
$\delta(q,a)$ is always represented by an NFA.

In our proofs, we will use reductions from the following decision
problems for string automata:
\begin{description}
\item[Emptiness:] Given an automaton $A$, is $L(A) = \emptyset$?
\item[Universality:] Given an automaton $A$, is $L(A) = \Sigma^*$?
\item[Intersection emptiness:] Given the automata $A_1, \ldots, A_n$,
  is $L(A_1) \cap \cdots \cap L(A_n) = \emptyset$?
\end{description}
The corresponding decision problems for tree automata are defined
analogously.  

In the Appendix, we show that the following statements hold over the
alphabet $\{0,1\}$ (Corollary~\ref{lem:tool-fix-alph}):
\begin{enumerate}
\item Intersection emptiness of an arbitrary number of DFAs is
  \pspace-hard.
\item Universality of NFAs  is \pspace-hard.
\end{enumerate}
Over the alphabet $\{0,1,0',1'\}$, the following statement holds:
\begin{enumerate}
\item[(3)] Intersection emptiness of an arbitrary number of TDBTAs is
  \exptime-hard.
\end{enumerate}

\subsection{Transducers}\label{sec:transducers}
We adhere to transducers as a formal model for simple transformations
corresponding to structural recursion~\cite{suciuvldbunql} and a
fragment of top-down XSLT.  As in \cite{msvjournal}, the abstraction
focuses on structure rather than on content. We next define the tree
transducers used in this paper.  To simplify notation, we restrict
ourselves to one alphabet.  That is, we consider transducers mapping
$\Sigma$-trees to $\Sigma$-trees.\footnote{In general, of course, one
  can define transducers where the input alphabet differs from the
  output alphabet.}

For a set $Q$, denote by $\cH_\Sigma(Q)$ (respectively
$\cT_\Sigma(Q)$) the set of $\Sigma$-hedges (respectively trees) where
leaf nodes are labeled with elements from $\Sigma \cup Q$ instead of
only $\Sigma$.

\begin{definition}
  A \emph{tree transducer} is a 4-tuple $T= (Q, \Sigma, q^0,R)$, where
  $Q$ is a finite set of states, $\Sigma$ is the input and output
  alphabet, $q^0\in Q$ is the initial state, and $R$ is a finite set
  of rules of the form $ (q,a) \to h$, where $a\in \Sigma$, $q\in Q$,
  and $h\in \cH_\Sigma(Q)$.  When $q=q^0$, $h$ is restricted to be
  either empty, or consist of only one tree with a $\Sigma$-symbol as
  its root label.
\end{definition}

The restriction on rules with the initial state ensures that the
output is always a tree rather than a hedge. Transducers are required
to be deterministic: for every pair $(q,a)$, there is at most one rule
in $R$.

The translation defined by a tree transducer $T = (Q,\Sigma,q^0,R)$ on
a tree $t$ in state $q$, denoted by $T^q(t)$, is inductively defined
as follows: if $t=\varepsilon$ then $T^q(t):=\varepsilon$; if
$t=a(t_1\cdots t_n)$ and there is a rule $(q,a)\to h\in R$ then
$T^q(t)$ is obtained from $h$ by replacing every node $u$ in $h$
labeled with state $p$ by the hedge $T^p(t_1)\cdots T^p(t_n)$. Note
that such nodes $u$ can only occur at leaves. So, $h$ is only extended
downwards.  If there is no rule $(q,a)\to h\in R$ then
$T^q(t):=\varepsilon$.  Finally, the transformation of $t$ by $T$,
denoted by $T(t)$, is defined as $T^{q^0}(t)$, interpreted as a tree.

For $a\in\Sigma$, $q\in Q$ and $(q,a)\to h\in R$, we denote $h$ by
rhs$(q,a)$. If $q$ and $a$ are not important, we say that $h$ is an
rhs. The \emph{size} of $T$ is $|Q|+ |\Sigma|+\sum_{q \in Q, a \in
  \Sigma} |\text{rhs}(q,a)|$, where $|\text{rhs}(q,a)|$ denotes the
number of nodes in $\text{rhs}(q,a)$. In the sequel, we always use $p,
p_1, p_2, \ldots$ and $q, q_1, q_2, \ldots$ to denote states.

Let $q$ be a state of tree transducer $T$ and $a\in\Sigma$. We then
define $q_T[a]:=\text{top}(T^q(a))$.  For a string $w = a_1\cdots
a_n$, we define $q_T[w]:=q_T[a_1]\cdots q_T[a_n]$. In the sequel, we
leave $T$ implicit whenever $T$ is clear from the context.

We give an example of a tree transducer:
\begin{example}\label{ex:tt}
  Let $T=(Q,\Sigma,p,R)$ where $Q=\{p,q\}$, $\Sigma = \{a, b,c,d,e\}$,
  and $R$ contains the rules $$
\begin{array}{lll}
  (p,a)\to d(e) &{}\quad{}& (p,b)\to d(q) \\
  (q,a)\to c\ p && (q,b) \to c(p\ q)
\end{array}
$$
Note that the right-hand side of $(q,a) \to c\ p$ is a hedge
consisting of two trees, while the other right-hand sides consist of
only one tree. 
\end{example}

Our tree transducers can be implemented as XSLT programs in a
straightforward way. For instance, the XSLT program equivalent to the
above transducer is given in Figure~\ref{fig:xslt} (we assume the
program is started in mode $p$). 

\begin{figure}[t]
\begin{center}
{\footnotesize
\begin{minipage}{7cm}
\begin{verbatim}
<xsl:template match="a" mode ="p">
  <d>
     <e/>
  </d>
</xsl:template>

<xsl:template match="b" mode ="p">
  <d>
     <xsl:apply-templates mode="q"/>
  </d>
</xsl:template>

<xsl:template match="a" mode ="q">
  <c/>
  <xsl:apply-templates mode="p"/>
</xsl:template>

<xsl:template match="b" mode ="q">
  <c>
      <xsl:apply-templates mode="p"/>
      <xsl:apply-templates mode="q"/>
  </c>
</xsl:template>
\end{verbatim}
\end{minipage}
}
\end{center}
\caption{\label{fig:xslt} The XSLT program equivalent to the transducer
of Example~\ref{ex:tt}.}
\end{figure}

\begin{figure*}[!tbh]
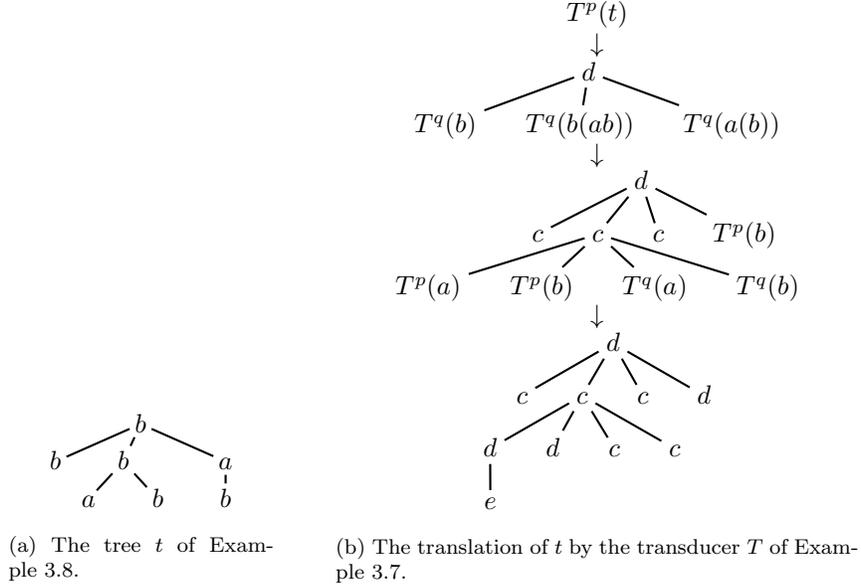

  \subfigure[The tree $t$ of Example~\ref{ex:t}.\label{fig:deriv-a}]{
     \begin{minipage}{4cm}
       \vspace{5.5cm}
       \begin{center}
         \pstree[nodesep=2pt,levelsep=0.5cm]{\TR{$b$}}{
           \TR{$b$}
           \pstree{\TR{$b$}}{\TR{$a$} \TR{$b$}}
           \pstree{\TR{$a$}}{\TR{$b$}}}
       \end{center}
     \end{minipage}
   }
  \subfigure[The translation of $t$ by the transducer $T$ of Example~\ref{ex:tt}.\label{fig:deriv-b}]{
    \begin{tabular}{c}
      \begin{minipage}{6cm}
        \begin{center}
          $T^{p}(t)$
        \end{center}
      \end{minipage}
      \\
      $\downarrow$
      \\
      \begin{minipage}{6cm}
        \begin{center}
          \pstree[treesep=0.65cm,levelsep=0.7cm,nodesep=0.1cm]{\TR{$d$}}{
            \TR{$T^q(b)$}
            \TR{$T^q(b(ab))$}
            \TR{$T^q(a(b))$}
          }
        \end{center}
      \end{minipage}
      \\
      $\downarrow$
      \\
      \begin{minipage}{6cm}
        \begin{center}
          \pstree[treesep=0.65cm,levelsep=0.7cm,nodesep=0.1cm]{\TR{$d$}}{
            \TR{$c$}
            \pstree{\TR{$c$}}{
              \TR{$T^p(a)$} 
              \TR{$T^p(b)$}
              \TR{$T^q(a)$} 
              \TR{$T^q(b)$}
            }
            \TR{$c$}
            \TR{$T^p(b)$}
          }
        \end{center}    
     \end{minipage}
      \\
      $\downarrow$ 
      \\
      \begin{minipage}{7cm}
        \begin{center}
          \pstree[treesep=0.65cm,levelsep=0.7cm,nodesep=0.1cm]{\TR{$d$}}{
            \TR{$c$}
            \pstree{\TR{$c$}}{
              \pstree{\TR{$d$}}{
                \TR{$e$}
              }
              \TR{$d$}
              \TR{$c$}
              \TR{$c$}
            }
            \TR{$c$}
            \TR{$d$}
          }
        \end{center}
      \end{minipage}
    \end{tabular}
}
  \caption{A tree and its translation.\label{fig:deriv}}
\end{figure*}

\begin{example}\rm\label{ex:t}
  Consider the tree $t$ shown in Figure~\ref{fig:deriv-a}.  In
  Figure~\ref{fig:deriv-b} we give the translation of $t$ by the
  transducer of Example~\ref{ex:tt}. In order to keep the example
  simple, we did not list $T^q(\varepsilon)$ and $T^p(\varepsilon)$
  explicitly in the process of translation.
\end{example}

We discuss two important features of tree transducers: \emph{copying}
and \emph{deletion}.  In Example~\ref{ex:tt}, the rule $(q,b)\to
c(p\,q)$ copies the children of the current node in the input tree
twice: one copy is processed in state $p$ and the other in state $q$.
The symbol $c$ is the parent node of the two copies. So, one could say
that the current node is translated in the new parent node labeled
$c$. The rule $(q,a)\to c\,p$ copies the children of the current node
only once.  However, no parent node is given for this copy. So, there
is no node in the output tree that can be interpreted as the
translation of the current node in the input tree.  We therefore say
that it is deleted.  For instance, $T^q(a(b))=c\,d$ where $d$
corresponds to $b$ and not to $a$.

We define some relevant classes of transducers.  A transducer is
\emph{non-deleting} if no states occur at the top-level of any rhs.
We denote by $\cT_{nd}$ the class of non-deleting transducers and by
$\cT_d$ the class of transducers where we allow deletion.
Furthermore, a transducer $T$ has \emph{copying width} $k$ if there
are at most $k$ occurrences of states in every sequence of siblings in
an rhs. For instance, the transducer in Example~\ref{ex:tt} has
copying width 2.  Given a natural number $k$, which we will leave
implicit, we denote by $\cT_{bc}$ the class of transducers of copying
width $k$. The abbreviation ``bc'' stands for \emph{bounded copying}.
We denote intersections of these classes by combining the indexes.
For instance, $\cT_{nd,bc}$ is the class of non-deleting transducers
with bounded copying. When we want to emphasize that we also allow
unbounded copying in a certain application, we write, for instance,
$\cT_{nd,uc}$ instead of $\cT_{nd}$.

\subsection{The Typechecking Problem}
\label{sec:tc:def}

\begin{definition}
A tree transducer $T$ \emph{typechecks} with respect to to an input
tree language $S_{\text{in}}$ and an output tree language $S_{\out}$,
if $T(t)\in S_{\out}$ for every $t\in S_{\text{in}}$.
\end{definition}

We now define the problem central to this paper.
\begin{definition}
  Given $S_{\text{in}}$, $S_{\out}$, and $T$, the \emph{typechecking
    problem} consists in verifying whether $T$ typechecks with respect
  to $S_{\text{in}}$ and $S_{\out}$.
\end{definition}

We parameterize the typechecking problem by the kind of tree
transducers and tree languages we allow. Let $\cT$ be a class of
transducers and ${\cS}$ be a representation of a class of tree
languages. Then TC[${\cT},{\cS}$] denotes the typechecking problem
where $T \in \cT$ and $S_{\text{in}},S_{\out}\in \cS$.  Examples of
classes of tree languages are those defined by tree automata or DTDs.
Classes of transducers are discussed in the previous section.  The
complexity of the problem is measured in terms of the sum of the sizes
of the input and output schemas $S_{\text{in}}$ and $S_{\out}$ and the
transducer $T$.

Table~\ref{tab:tc-complexities:old} summarizes the results obtained
in~\cite{martensneventcs05}.  Unless specified otherwise, all problems
are complete for the mentioned complexity classes.  In the setting
of~\cite{martensneventcs05}, typechecking is only tractable when
restricting to non-deleting and bounded copying transducers in the
presence of DTDs with DFAs. 

Recall that, in this article, we are interested in variants of the
typechecking problem where the input and/or output schema is fixed. We
therefore introduce some notations that are central to the paper.  We
denote the typechecking problem where the input schema, the output
schema, or both are fixed by TC$^i$[${\cT},{\cS}$],
TC$^o$[${\cT},{\cS}$], and TC$^{io}$[${\cT},{\cS}$], respectively. The
complexity of these subproblems is measured in terms of the sum of the
sizes of the input and output schemas $S_{\text{in}}$ and $S_{\out}$,
and the transducer $T$, minus the size of the fixed schema(s).

\begin{table}[bt]
\begin{center}
\begin{tabular}{|l||c|c|c|c|c|}
\hline
 & NTA & DTA& DTD(NFA)& DTD(DFA) & DTD(SL)\\
\hline \hline
d,uc & \exptime & \exptime & \exptime & \exptime & \exptime\\
\hline 
nd,uc & \exptime & \exptime & \pspace & \pspace & \conp\\
\hline 
nd,bc
& \exptime & 
\renewcommand{\arraystretch}{0.6}
\begin{tabular}{c}
  in \exptime \\ 
  \pspace-hard
\end{tabular} 
\renewcommand{\arraystretch}{1}
& \pspace & \ptime & \conp\\
\hline
\end{tabular}
\end{center}
\caption{Results of \protect\cite{martensneventcs05} 
  (upper and lower bounds). The top row shows the representation of the input and output schemas and the
  left column shows the class of tree transducer: ``d'', ``nd'', ``uc'',
  and ``bc'' stand for ``deleting'', ``non-deleting'', ``unbounded
  copying'', and ``bounded copying'' respectively.\label{tab:tc-complexities:old}}
\end{table}

\section{Main Results}
\label{sec:fixed}

\begin{table*}[htb]
\begin{center}
\begin{tabular}{|l|l||c|c|c|c|c|}
  \hline
  fixed & TT & NTA & DTA & DTD(NFA) & DTD(DFA) & DTD(SL)\\
  \hline \hline
  in, out, & d,uc & \bexptime & \bexptime & \bexptime & \bexptime & \bexptime\\
  \cline{2-7}
  in+out   & d,bc & \bexptime & \bexptime & \bexptime & \bexptime & \bexptime\\
  \hline \hline
  in  & nd,uc & \bexptime & \bexptime & \bpspace & \bpspace &
  \underline{in \bptime} \\
      \cline{2-7}
      & nd,bc & \bexptime & \bexptime & \bpspace  &
      \underline{\bnlogspace} & \underline{in \bptime} \\
  \hline
  out & nd,uc & \bexptime & \bexptime & \bpspace & \bpspace & \conp \\
      \cline{2-7}
      & nd,bc & \bexptime & \bexptime & \underline{\bptime} & \ptime &\conp\\
  \hline
  in+out & nd,uc & \bexptime & \bexptime & \underline{\bnlogspace} &
  \underline{\bnlogspace} & \underline{\bnlogspace}  \\
         \cline{2-7}
         & nd,bc & \bexptime & \bexptime & \underline{\bnlogspace} &
  \underline{\bnlogspace} & \underline{\bnlogspace}\\
  \hline
\end{tabular}
\caption{Complexities of the typechecking
  problem in the new setting (upper and lower bounds). The
  top row shows the representation of the input and output
  schemas, the leftmost column shows which schemas are fixed, and the
  second column to the left shows the class of tree transducer: ``d'',
  ``nd'', ``uc'', and ``bc'' stand for ``deleting'', ``non-deleting'',
  ``unbounded copying'', and ``bounded copying'' respectively. In the
  case of deleting transformations, the different possibilities are
  grouped as all complexities coincide. 
  \label{tab:tc-complexities}}
\end{center}
\end{table*}

As argued in the Introduction, it makes sense to consider the input
and/or output schema not as part of the input for some scenarios. From
a complexity theory point of view, it is important to note here that
the input and/or output alphabet then also becomes fixed.  In this
article, we revisit the results
of~\cite{martensneventcs05} from that perspective.

The results are summarized in Table~\ref{tab:tc-complexities}.  As
some results already follow from proofs in~\cite{martensneventcs05},
we printed the results requiring a new proof in bold.  The entries
where the complexity was lowered (assuming that the complexity classes
in question are different) are underlined. Again, all problems are
complete for the mentioned complexity classes unless specified
otherwise.

We discuss the obtained results: for non-deleting transformations, we
get three new tractable cases: \emph{(i)} fixed input schema,
\emph{un}bounded copying, and DTD(SL)s; \emph{(ii)} fixed output schema,
bounded copying and DTD(NFA)s; and, \emph{(iii)} fixed input and output,
\emph{un}bounded copying and all DTDs.  It is striking, however, that
in the presence of deletion or tree automata (even deterministic ones)
typechecking remains \exptime-hard for \emph{all} scenarios.

Mostly, we only needed to strengthen the lower bound proofs
of~\cite{martensneventcs05}.

\subsection{Deletion: Fixed Input Schema, Fixed Output Schema, and
  Fixed Input and Output Schema}

The \exptime upper bound for typechecking already follows
from~\cite{martensneventcs05}. Therefore, it remains to show the lower
bounds for TC$^{io}$[$\cT_{d,bc}$,DTD(DFA)] and
TC$^{io}$[$\cT_{d,bc}$,DTD(SL)], which we do in
Theorem~\ref{theo:fix-io-dw2-bc-exptime}.  In fact, if follows from
the proof that the lower bounds already hold for transducers with
copying width 2.

We require the notion of top-down deterministic binary tree automata
in the proof of Theorem~\ref{theo:fix-io-dw2-bc-exptime}. A
\emph{binary tree automaton} (BTA) is a non-deterministic tree
automaton $B = (Q,\Sigma,\delta,F)$ operating on \emph{binary trees}.
These are trees where every node has zero, one, or two children. We
assume that the alphabet is partitioned in \emph{internal labels} and
\emph{leaf labels}. When a label $a$ is an internal label, the regular
language $\delta(q,a)$ only contains strings of length one or two.
When $a$ is a leaf label, the regular language $\delta(q,a)$ only
contains the empty string.  A binary tree automaton is
\emph{top-down deterministic} if \emph{(i)} $F$ is a singleton and,
\emph{(ii)} for every $q,q'\in Q$ with $q\neq q'$ and $a\in\Sigma$,
$\delta(q,a)$ contains at most one string. We abbreviate ``top-down
deterministic binary tree automaton'' by TDBTA.

\begin{theorem}\label{theo:fix-io-dw2-bc-exptime}
  \begin{enumerate}
  \item $TC^{io}[\cT_{d,bc},DTD(DFA)]$ is \exptime-complete; and
  \item $TC^{io}[\cT_{d,bc},DTD(SL)]$ is \exptime-complete.
  \end{enumerate}
\end{theorem}
\begin{proof}
  The \exptime upper bound follows from Theorem~11
  in~\cite{martensneventcs05}. We proceed by proving the lower bounds.

  We give a \logspace reduction from the intersection emptiness
  problem of an arbitrary number of top-down deterministic binary tree
  automata (TDBTAs) over the alphabet $\Sigma = \{0,1,0',1'\}$. The
  intersection emptiness problem of TDBTAs over alphabet
  $\{0,1,0',1'\}$ is known to be \exptime-hard (\mbox{cfr.}
  Corollary~\ref{lem:tool-fix-alph}(3) in the Appendix).
  
  For $i = 1, \ldots, n$, let $A_i = (Q_i, \Sigma, \delta_i,
  \{\text{start}_i\})$ be a TDBTA, with $\Sigma = \{0,1,0', 1'\}$.
  Without loss of generality, we can assume that the state sets $Q_i$
  are pairwise disjoint. We call $0$ and $1$ \emph{internal labels}
  and $0'$ and $1'$ \emph{leaf labels}.  In our proof, we use the
  markers `$\ell$' and `$r$' to denote that a certain node is a left
  or a right child.  Formally, define $\Sigma_\ell := \{a_\ell \mid a
  \in \Sigma\}$ and $\Sigma_r := \{a_r \mid a \in \Sigma\}$. We use
  symbols from $\Sigma_\ell$ and $\Sigma_r$ for the left and right
  children of nodes, respectively.

  We now define a transducer $T$ and two DTDs $d_{\text{in}}$ and
  $d_{\text{out}}$ such that $\bigcap_{i=1}^n L(A_i) = \emptyset$ if
  and only if $T$ typechecks with respect to $d_{\text{in}}$ and
  $d_{\text{out}}$.  In the construction, we exploit the copying power
  of transducers to make $n$ copies of the input tree: one for each
  $A_i$. By using deleting states, we can execute each $A_i$ on its
  copy of the input tree without producing output.  When an $A_i$ does
  not accept, we output an \emph{error} symbol under the root of the
  output tree. The output DTD should then only check that an
  \emph{error} symbol always appears. A bit of care needs to be taken,
  as a bounded copying transducer can not make an arbitrary number of
  copies of the input tree in the same rule. The transducer therefore
  goes through an initial copying phase where it repeatedly copies
  part of the input tree twice, until there are (at least) $n$
  copies. The transducer remains in the copying phase as long as it
  processes special symbols ``$\#$''. The input trees are therefore of
  the form as depicted in Figure~\ref{fig:treeform}. In addition, the
  transducer should verify that the number of $\#$-symbols in the
  input equals $\lceil \log n \rceil$.

  \begin{figure}
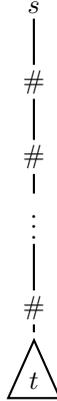

  \begin{center}
    \pstree[nodesep=2pt,levelsep=1cm,treesep=0.5cm]{\TR{$s$}}{
      \pstree{\TR{$\#$}}{
        \pstree{\TR{$\#$}}{
          \pstree{\TR{$\vdots$}}{
            \pstree{\TR{$\#$}}{
              \TR{\trinode{X}{$t$}}
            }
          }
        }
      }
    }
  \end{center}
  \caption{Structure of the trees defined by the input schema in the
    proof of Theorem~\ref{theo:fix-io-dw2-bc-exptime}.
    \label{fig:treeform}}
  \end{figure}

  The input DTD $(d_\text{in},s)$, which we will describe next, uses
  the alphabet $\Sigma_\ell \cup \Sigma_r \cup \{s,\#\}$, and defines
  all trees of the form as described in Figure~\ref{fig:treeform},
  where $s$ and $\#$ are alphabet symbols, and every internal node of
  $t$ (which is depicted in Figure~\ref{fig:treeform}) has one or two
  children. When a node is an only child, it is labeled with an
  element of $\Sigma_\ell$. Otherwise, it is labeled with an element
  of $\Sigma_\ell$ or an element of $\Sigma_r$ if it is a left child
  or a right child, respectively. In this way, the transducer knows
  whether a node is a left or a right child by examining the label.
  The root symbol of $t$ is labeled with a symbol from $\Sigma_\ell$.
  Furthermore, all internal nodes of $t$ are labeled with labels in
  $\{0_\ell,0_r,1_\ell,1_r\}$ and all leaf nodes are labeled with
  labels in $\{0'_\ell,0'_r,1'_\ell,1'_r\}$.  As explained above, we
  will use the sequence of $\#$-symbols to make a sufficient number of
  copies of $t$.
  
  The input DTD $(d_\text{in},s)$ is defined as follows:
  \begin{itemize}
  \item $d_\text{in}(s) = \# + 0_\ell + 1_\ell$;
  \item $d_\text{in}(\#) = \# + 0_\ell + 1_\ell$;
  \item for each $a \in \{0_\ell,1_\ell,0_r,1_r\}$,\\ $d_\text{in}(a) =
    (0_\ell + 1_\ell + 0'_\ell + 1'_\ell) + (0_\ell + 1_\ell + 0'_\ell
    + 1'_\ell)(0_r + 1_r + 0'_r + 1'_r)$; and,
  \item for each $a \in \{0'_\ell,1'_\ell, 0'_r, 1'_r\}$,
    $d_\text{in}(a) = \varepsilon$.
  \end{itemize}
  Obviously, $(d_\text{in},s)$ can be expressed as a DTD(DFA). It can
  also be expressed as a DTD(SL), as follows 
  \begin{align*}
    d_\text{in}(a) =  \bigg( & \big(
    (\varphi[0_\ell^{=1}] \lor \varphi[1_\ell^{=1}] \lor \varphi[(0'_\ell)^{=1}] \lor
    \varphi[(1'_\ell)^{=1}]) \big) \\  \oplus & \big( (\varphi[0_\ell^{=1}] \lor
    \varphi[1_\ell^{=1}] \lor \varphi[(0'_\ell)^{=1}] \lor
    \varphi[(1'_\ell)^{=1}])\\ & \land
     (\varphi[0_r^{=1}] \lor \varphi[1_r^{=1}] \lor \varphi[(0'_r)^{=1}] \lor
    \varphi[(1'_r)^{=1}])\big) \bigg) \\ & \land s^{=0} \land \#^{=0}
  \end{align*}
  for every $a \in \{0_\ell,1_\ell,0_r,1_r\}$, where 
  \begin{itemize}
  \item $\oplus$ denotes the ``exclusive or''; 
  \item for every $i \in \{\ell,r\}$ and $x \in \{0_i,1_i,0'_i,1'_i\}$,
    $\varphi[x^{=1}]$ denotes the conjunction $$(x^{=1}
    \land \bigwedge_{y \in \{0_i,1_i,0'_i,1'_i\} \setminus
      \{x\}} y^{=0}).$$
  \end{itemize}
  Notice that the size of the SL-formula expressing $d_\text{in}(a)$
  is constant.
  
  We construct a tree transducer $T = (Q_T, \Sigma_T,
  q^\varepsilon_\text{copy}, R_T)$.  The alphabet of $T$ is $\Sigma_T
  = \Sigma_\ell \cup \Sigma_r \cup \{s, \#, \text{error},
  \text{ok}\}$.  The state set $Q_T$ is defined to be the set
  $\{q^\ell, q^r \mid q \in Q_i, i \in \{1,\ldots,n\}\}$.  The
  transducer will use $\lceil \log n \rceil$ special copying states
  $q_\text{copy}^j$ to make at least $n$ copies of the input tree. To
  define $Q_T$ formally, we first introduce the notation $D(k)$, for
  $k = 0,\ldots,\lceil \log n \rceil$. Intuitively, $D(k)$ corresponds
  to the set of nodes of a complete binary tree of depth $k+1$. For
  example, $D(1) = \{\varepsilon,0,1\}$ and $D(2) =
  \{\varepsilon,0,1,00,01,10,11\}$. The idea is that, if $i \in D(k)
  \setminus D(k-1)$, for $k > 0$, then $i$ represents the binary
  encoding of a number in $\{0,\ldots,2^k-1\}$. Formally, if $k = 0$,
  then $D(k) = \{\varepsilon\}$; otherwise, $D(k) = D(k-1) \cup
  \bigcup_{j = 0,1} \{ij \mid i \in D(k-1)\}$.  The state set $Q_T$ is
  then the union of the sets $Q^\ell = \{q^\ell \mid q \in Q_j, 1 \leq
  j \leq n\}$, $Q^r = \{q^r \mid q \in Q_j, 1 \leq j \leq n\}$, the
  set $\{q^j_\text{copy} \mid j \in D(\lceil \log n \rceil)\}$ and the
  set $\{\text{start}_j^\ell \mid n+1 \leq j \leq 2^{\lceil \log n
    \rceil}\}$. Note that the last set can be empty. It only contains
  dummy states translating any input to the empty string.
  
  We next describe the action of the tree transducer $T$. Roughly, the
  operation of $T$ on the input $s(\#(\#(\cdots\#(t))))$ can be
  divided in two parts: \emph{(i)} copying the tree $t$ a sufficient
  number of times while reading the $\#$-symbols; and, \emph{(ii)}
  simulating one of the TDBTAs on each copy of $t$. The tree
  transducer outputs the symbol ``error'' when one of the TDBTAs
  rejects $t$, or when the number of $\#$-symbols in its input is not
  equal to $\lceil \log n \rceil$.  Apart from copying the root symbol
  $s$ to the output tree, $T$ only writes the symbol ``error'' to the
  output. Hence, the output tree always has a root labeled $s$ which
  has zero or more children labeled ``error''. The output DTD, which
  we define later, should then verify whether the root has always one
  ``error''-labeled child.

  Formally, the transition rules in $R_T$ are defined as follows:
  \begin{itemize}
  \item $(q^\varepsilon_\text{copy},s) \to s(q^0_\text{copy}
    q^1_\text{copy})$. This rule puts $s$ as the root symbol of the
    output tree.
  \item $(q^i_\text{copy},\#) \to q^{i0}_\text{copy}
    q^{i1}_\text{copy}$ for $i \in D(\lceil \log n \rceil - 1) -
    \{\varepsilon\}$. These rules copy the tree $t$ in the input at
    least $n$ times, provided that there are enough $\#$-symbols.
  \item $(q^i_\text{copy},\#) \to \text{start}^\ell_k$, where $i \in
    D(\lceil \log n \rceil) - D(\lceil \log n \rceil -1)$, and $i$ is
    the binary representation of $k$.  This rule starts the
    in-parallel simulation of the $A_i$'s. For $i = n+1, \ldots,
    2^{\lceil \log n \rceil}$, $\text{start}^\ell_i$ is just a dummy
    state transforming everything to the empty tree.
  \item $(q^i_\text{copy},a) \to \text{error}$ for $a \in \Sigma$ and
    $i \in D(\lceil \log n \rceil)$. This rule makes sure that the
    output of $T$ is accepted by the output tree automaton if there
    are not enough $\#$-symbols in the input.
  \item $(\text{start}^\ell_k,\#) \to \text{error}$ for all $k =
    1,\ldots,2^{\lceil \log n \rceil}$. This rule makes sure that the
    output of $T$ is accepted by the output tree automaton if there
    are too much $\#$-symbols in the input.
  \item $(q^\ell,a_r) \to \varepsilon$ and $(q^r,a_\ell) \to
    \varepsilon$ for all $q \in Q_j$, $j = 1,\ldots,n$. This rule
    ensures that tree automata states intended for left (respectively
    right) children are not applied to right (respectively left)
    children.
  \item $(q^\ell,a_\ell) \to q_1^\ell q_2^r$ and $(q^r,a_r) \to
    q_1^\ell q_2^r$, for every $q \in Q_i$, $i = 1,\ldots,n$, such
    that $\delta_i(q,a) = q_1q_2$, and $a$ is an internal symbol. This
    rule does the actual simulation of the tree automata $A_i$, $i =
    1,\ldots,n$.
  \item $(q^\ell,a_\ell) \to q_1^\ell$ and $(q^r,a_r) \to q_1^\ell$,
    for every $q \in Q_i$, $i = 1,\ldots,n$, such that $\delta_i(q,a)
    = q_1$ and $a$ is an internal symbol. This rule does the actual
    simulation of the tree automata $A_i$, $i = 1,\ldots,n$.
  \item $(q^\ell,a_\ell) \to \varepsilon$ and $(q^r,a_r) \to
    \varepsilon$ for every $q \in Q_i$, $i = 1,\ldots,n$, such that
    $\delta_i(q,a) = \varepsilon$ and $a$ is a leaf symbol. This rule
    simulates accepting computations of the $A_i$'s.
  \item $(q^\ell,a_\ell) \to $ error and $(q^r,a_r) \to$ error for
    every $q \in Q_i$, $i = 1,\ldots,n$, such that $\delta_i(q,a)$ is
    undefined. This rule simulates rejecting computations of the
    $A_i$'s.
  \end{itemize}
  It is straightforward to verify that, on input
  $s(\#(\#(\cdots\#(t))))$, $T$ outputs the tree $s$ if and only if
  there are $\lceil \log n \rceil$ $\#$-symbols in the input and $t
  \in L(A_1) \cap \cdots \cap L(A_n)$.
  
  Finally, $d_\out(s) = \text{error} \ \text{error}^*$, which can
  easily be defined as a DTD(DFA) and as a DTD(SL).

  It is easy to see that the reduction can be carried out in
  deterministic logarithmic space, that $T$ has copying width 2, and
  that $d_\text{in}$ and $d_\text{out}$ do not depend on
  $A_1,\ldots,A_n$. \hfill $\Box$
\end{proof}

\subsection{Non-deleting: Fixed Input Schema}

We turn to the typechecking problem in which we consider the input
schema as fixed. We start by showing that typechecking is in \ptime in
the case where we use DTDs with SL-expressions and the tree transducer
is non-deleting (Theorem~\ref{theo:nd,c,sl}). To this end, we recall a
lemma and introduce some necessary notions that are needed for the
proof of Theorem~\ref{theo:nd,c,sl}.

For an SL-formula $\phi$, we say that two strings $w_1$ and $w_2$ are
\emph{$\phi$-equivalent} (denoted $w_1 \equiv_\phi w_2$) if $w_1
\models \phi$ if and only if $w_2 \models \phi$.  

For $a\in\Sigma$ and $w\in\Sigma^*$, we denote by $\#_a(w)$ the number
of $a$'s occurring in $w$. We recall Lemma 17
from~\cite{martensneventcs05}:
\begin{lemma}\label{lem:SL-equiv}
  Let $\phi$ be an SL-formula and let $k$ be the largest integer
  occurring in $\phi$. For every $w,w' \in \Sigma^*$, for every $a \in
  \Sigma$,
  \begin{itemize}
  \item if $\#_a(w') > k$ when $\#_a(w) > k$, and
  \item $\#_a(w') = \#_a(w)$, otherwise, 
  \end{itemize}
  then $w \equiv_\phi w'$.
\end{lemma}

For a hedge $h$ and a DTD $d$, we say that $h$ \emph{partly satisfies}
$d$ if for every $u\in\Dom(h)$, $\lab^h(u1)\cdots\lab^h(un)\in
L(d(\lab^h(u)))$ where $u$ has $n$ children. Note that there is no
requirement on the root nodes of the trees in $h$. Hence, the term
``partly''.

We are now ready to show the first \ptime result:
\begin{theorem} \label{theo:nd,c,sl}
  $TC^i[\cT_{nd,uc}$, DTD(SL)$]$ is in \ptime.
\end{theorem}
\begin{proof}
  Denote the tree transformation by $T = (Q_T, \Sigma, q^0_T, R_T)$
  and the input and output DTDs by $(d_\text{in},s_\text{in})$ and
  $(d_\out,s_\out)$, respectively. As $d_\text{in}$ is fixed, we can
  assume that $d_\text{in}$ is reduced.

  Intuitively, the typechecking algorithm is successful when $T$ does
  \emph{not} typecheck with respect to $d_\text{in}$ and
  $d_\text{out}$. The outline of the typechecking algorithm is as
  follows:
  \begin{enumerate}
  \item Compute the set $RP$ of ``reachable pairs'' $(q,a)$ for which
    there exists a tree $t \in L(d_\text{in})$ and a node $u \in
    \Dom(t)$ such that $\lab^t(u) = a$ and $T$ visits $u$ in state
    $q$. That is, we compute all pairs $(q,a)$ such that either
    \begin{itemize}
    \item $q = q^0_T$ and $a = s_\text{in}$; or
    \item $(q',a') \in RP$, there is a $q$-labeled node in
      $\rhs(q',a')$, and there exists a string $w_1aw_2 \in
      d_\text{in}(a')$ for $w_1, w_2 \in \Sigma^*$.
    \end{itemize}
    
  \item For each such pair $(q,a)$ and for each node $v \in
    \Dom(\rhs(q,a))$, test whether there exists a string $w \in
    d_\text{in}(a)$ such that $T^q(a(w))$ does not partly satisfy
    $d_\out$. We call $w$ a \emph{counterexample}.
  \end{enumerate}
  The algorithm is successful, if and only if there exists a
  counterexample.

  We illustrate the general operation of the typechecking algorithm in
  Figure~\ref{fig:tcguess}. In this figure, $T$ visits the $a$-labeled
  node on the left in state $q$. Consequently, $T$ outputs the hedge
  $\rhs(q,a)$, which is illustraded by dotted lines on the right. The
  typechecking algorithm searches for a node $u$ in $\rhs(q,a)$ (which
  is labeled by $c$ in the figure), such that the string of children
  of $u$ is not in $L(d_\text{out}(c))$.

  \begin{figure}
    \begin{center}
      \resizebox{!}{7cm}{\input{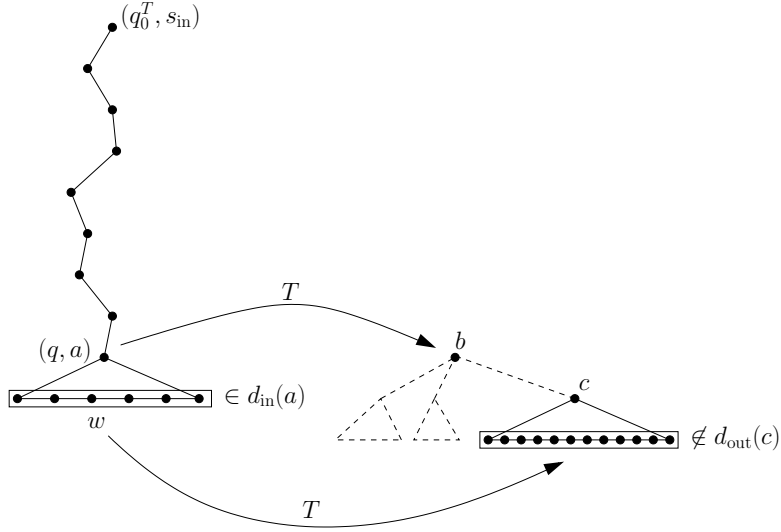}}
    \end{center}
    \caption{Illustration of the typechecking algorithm in the proof
      of Theorem~\ref{theo:nd,c,sl}. \label{fig:tcguess}}
  \end{figure}
  
  Notice that the typechecking algorithm does not assume that $d_\out$
  is reduced (recall the definition of a reduced DTD from
  Section~\ref{sec:DTD-TA}).  We need to show that the algorithm is
  correct, that is, there exists a counterexample if and only if $T$
  does not typecheck with respect to $d_\text{in}$ and $d_\text{out}$.
  Clearly, when the algorithm does not find a counterexample, $T$
  typechecks with respect to $d_\text{in}$ and $d_\text{out}$.
  Conversely, suppose that the algorithm finds a pair $(q,a)$ and a
  string $w$ such that $T^q(a(w))$ does not partly satisfy $d_\out$.
  So, since we assumed that $d_\text{in}$ is reduced, there exists a
  tree $t \in L(d_\text{in})$ and a node $u \in \Dom(t)$ such that
  $\lab^t(u) = a$ and $u$ is visited by $T$ in state $q$. Also, there
  exists a node $v$ in $T^q(a(w))$, such that the label of $u$ is $c$
  and the string of children of $u$ is not in $d_\text{out}(c)$. We
  argue that $T(t) \not \in L(d_\text{out})$.  There are two cases:
  \begin{enumerate}[(i)]
  \item if $L(d_\text{out})$ contains a tree with a $c$-labeled node,
    then $T(t) \not \in d_\out$ since $T^q(a(w))$ does not partly
    satisfy $d_\out$; and
  \item if $L(d_\text{out})$ does \emph{not} contain a tree with a
    $c$-labeled node, then $T(t) \not \in d_\out$ since $T(t)$
    contains a $c$-labeled node.
  \end{enumerate}

  We proceed by showing that the algorithm can be carried out in
  polynomial time.  As the input schema is fixed, step (1) of the
  algorithm is in polynomial time. Indeed, we can compute the set $RP$
  of reachable pairs $(q,a)$ in a top-down manner by a straightforward
  reachability algorithm.

  To show that step (2) of the typechecking algorithm is in polynomial
  time, fix a tuple $(q,a)$ that was computed in step (1) and a node
  $u$ in $\rhs(q,a)$ with label $b$. Let $z_0 q_1 z_1 \cdots q_n z_n$
  be the concatenation of $u$'s children, where all $z_0,\ldots,z_n
  \in \Sigma^*$ and $q_1,\ldots,q_n \in Q_T$. We now search for a
  string $w \in \Sigma^*$ for which $w \models d_\text{in}(a)$, but
  for which $z_0 q_1[w] z_1 \cdots q_n[w] z_n \not \models d_\out(b)$.
  Recall from Section~\ref{sec:transducers} that $q[w]$ is the
  homomorphic extension of $q[a]$ for $a \in \Sigma$, which is
  $\text{top}(\text{rhs}(q,a)))$ in the case of non-deleting tree
  transducers.

  Denote $d_\text{in}(a)$ by $\phi$. Let $\{a_1,\ldots,a_s\}$ be the
  different symbols occurring in $\phi$ and let $k$ be the largest
  integer occurring in $\phi$.  According to Lemma~\ref{lem:SL-equiv},
  every $\Sigma$-string is $\phi$-equivalent to a string of the form
  $w= a_1^{m_1} \cdots a_s^{m_s}$ with $0 \leq m_i \leq k+1$ for each
  $i=1,\ldots,s$.  Note that there are $(k+1)^s$ such strings, which
  is a constant number, as it only depends on the input schema.
  For the following, the algorithm considers each
  such string $w$.

  Fix such a string $w$ such that $w \models \phi$.  For each symbol
  $c$ in $d_\out(b)$, the number $\#_c(z_0 q_1[w] z_1 \cdots q_n[w]
  z_n)$ is equal to the linear sum $$k_1^c \times \#_{a_1}(w) + \cdots
  + k_\ell^c \times \#_{a_\ell}(w) + k_{\ell+1}^c \times
  \#_{a_{\ell+1}}(w) + k_s^c \times \#_{a_s}(w) + k^c,$$
  where $k^c =
  \#_c(z_0 \cdots z_n)$ and for each $i = 1,\ldots,s$, $k_i^c =
  \#_c(q_1[a_i] \cdots q_n[a_i])$. We now must test if there exists a
  string $w' \equiv_\phi w$ such that $z_0 q_1[w'] z_1 \cdots q_n[w']
  z_n \not \models d_\out(b)$.  Let $a_1, \ldots, a_{\ell}$ be the
  symbols that occur at least $k+1$ times in $w$ and $a_{\ell+1},
  \ldots, a_s$ be the symbols that occur at most $k$ times in $w$,
  respectively. Then, deciding whether $w'$ exists is equivalent to
  finding an integer solution to the variables
  $x_{a_1},\ldots,x_{a_s}$ for the boolean combination of linear
  (in)equalities $\Phi = \Phi_1 \land \lnot \Phi_2$, where

  \begin{itemize}
  \item $\Phi_1$ states that $w' \equiv_\phi w$, that is, 
    $$\Phi_1 = \bigwedge_{i=1}^\ell(x_{a_i} > k) \land
    \bigwedge_{j=\ell+1}^s\big(x_{a_j} = \#_{a_j}(w)\big);$$ and
  \item $\Phi_2$ states that $z_0 q_1(w') z_1 \cdots q_n(w') z_n
    \models d_\out(b)$, that is, $\Phi_2$ is defined by replacing
    every occurrence of $c^{=i}$ or $c^{\geq i}$ in $d_\out(b)$ by the
    equation $$\sum_{j=1}^s (k_j^c \times x_{a_j}) + k^c = i$$
    or by
    $$\sum_{j=1}^s (k_j^c \times x_{a_j}) + k^c \geq i,$$
    respectively.
  \end{itemize}
  In the above (in)equalities, $x_{a_i}$, $1 \leq i \leq s$,
  represents the number of occurrences of $a_i$ in $w'$.
  
  Finding a solution for $\Phi$ now consists of finding integer values
  for $x_{a_1}, \ldots, x_{a_s}$ so that $\Phi$ evaluates to
  \texttt{true}. Corollary~\ref{cor:IP} in the Appendix shows that we
  can decide in \ptime whether such a solution for $\Phi$ exists.
  \hfill $\Box$
\end{proof}

\begin{theorem}\label{theo:fix-i-nd-bc-dfa}
  $TC^i[\cT_{nd,bc}$, DTD(DFA)$]$ is \nlogspace-complete.
\end{theorem}
\begin{proof}
  In Theorem~\ref{theo:fix-io-nd-bc-nl-hard}(2), we prove that the
  problem is \nlogspace-hard, even if both the input and output
  schemas are fixed. Hence, it remains to show that the problem is in
  \nlogspace.

  Let us denote the tree transformation by $T = (Q_T, \Sigma, q^0_T,
  R_T)$ and the input and output DTDs by $(d_\text{in},r)$ and
  $d_\out$, respectively. We can assume that $d_\text{in}$ is
  reduced.\footnote{Reducing $d_\text{in}$ would be \ptime-complete
    otherwise, see Corollary~\ref{cor:reducing} in the Appendix.}

Then, the typechecking algorithm can be summarized as
  follows:
  \begin{enumerate}
  \item Guess a sequence of pairs $(q_0,a_0),(q_1,a_1), \ldots,
    (q_n,a_n)$ in $Q_T \times \Sigma_\text{in}$, such that 
    \begin{itemize}
    \item $(q_0,a_0) = (q^0_T,r)$; and
    \item for every pair $(q_i,a_i)$, $q_{i+1}$ occurs in
      rhs$(q_i,a_i)$ and $a_{i+1}$ occurs in some string in
      $L(d_\text{in}(a_i))$.
    \end{itemize}
    We only need to remember $(q_n,a_n)$ as a result of this step.
  \item Guess a node $u$ in rhs$(q_n,a_n)$ --- say that $u$ is labeled
    with $b$ --- and test whether there exists a string $w \in
    d_\text{in}(a_n)$ such that $T^q(a_n(w))$ does not partly satisfy
    $d_\out$.
  \end{enumerate}
  The algorithm is successful if and only if $w$ exists and, hence,
  the problem does not typecheck.
  
  The first step is a straightforward reachability algorithm, which is
  in \nlogspace.  It remains to show that the second step is in
  \nlogspace.
  
  Let $(q,a)$ be the pair $(q_n,a_n)$ computed in step two.  Let
  $d_\out(b) = (Q_\out, \Sigma, \delta_\out, \allowbreak \{p_I\},
  \{p_F\})$ be a DFA and let $k$ be the copying bound of $T$.  Let
  $z_0 q_1 z_1 \cdots q_{\ell} z_{\ell}$ be the concatenation of $u$'s
  children, where $\ell \leq k$. So we want to check whether there
  exists a string $w$ such that $z_0 q_1[w] z_1 \cdots q_{\ell}[w]
  z_{\ell}$ is not accepted by $d_\out(b)$. We guess $w$ one symbol at
  a time and simulate in parallel $\ell$ copies of $d_\out(b)$ and one
  copy of $d_\text{in}(a)$.

  By $\hat \delta$ we denote the canonical extension of $\delta$ to
  strings in $\Sigma^*$. We start by guessing states $p_1, \ldots,
  p_{\ell}$ of $d_\out(b)$, where $p_1 = \hat \delta_\out(p_I,z_0)$,
  and keep a copy of these on tape, to which we refer as $p'_1,
  \ldots, p'_{\ell}$.  Next, we keep on guessing symbols $c$ of $w$,
  whereafter we replace each $p_i$ by $\hat \delta_\out(p_i,q_i(c))$.
  The input automaton obviously starts in its initial state and is
  simulated in the straightforward way.

  The machine non-deterministically stops guessing, and checks
  whether, for each $i = 1, \ldots, \ell-1$, $\hat \delta_\out(p_i,
  z_i) = p'_{i+1}$ and $\hat \delta_\out(p_{\ell},z_{\ell}) = p_F$.
  For the input automaton, it simply checks whether the current state
  is the final state. If the latter tests are positive, then the
  algorithm accepts, otherwise, it rejects.

  We only keep $2 \ell + 1$ states on tape, which
  is a constant number, so the algorithm runs in \nlogspace. \hfill
  $\Box$
\end{proof}

\begin{theorem}
\begin{enumerate}
\item $TC^i[\cT_{nd,uc}$, DTD(DFA)$]$ is \pspace-complete; and
\item $TC^i[\cT_{nd,bc}$, DTD(NFA)$]$ is \pspace-complete.
\end{enumerate}
\end{theorem}
\begin{proof}
  In~\cite{martensneventcs05}, it was shown that both problems are in
  \pspace. We proceed by showing that they are also \pspace-hard.  

  (1) We reduce the intersection emptiness problem of an arbitrary
  number of deterministic finite automata with alphabet $\{0,1\}$ to
  the typechecking problem. This problem is known to be \pspace-hard,
  as shown in Corollary~\ref{lem:tool-fix-alph}(1) in the Appendix.
  Our reduction only requires logarithmic space. We define a
  transducer $T = (Q_T, \{0,1,\#_0, \ldots, \#_n\}, q^0_T, R_T)$ and
  two DTDs $d_\text{in}$ and $d_\text{out}$ such that $T$ typechecks
  with respect to $d_\text{in}$ and $d_\text{out}$ if and only if
  $\bigcap_{i=1}^n L(M_i) = \emptyset$.

  The DTD $(d_\text{in},s)$ defines trees of depth two, where the
  string formed by the children of the root is an arbitrary string in
  $\{0,1\}^*$, so $d_\text{in}(s) = (0+1)^*$. The transducer makes $n$
  copies of this string, separated by the delimiters $\#_i$:
  $Q_T=\{q,q^0_T\}$ and $R_T$ contains the rules $(q^0_T,s) \to s(\#_0
  q \#_1 q \ldots \#_{n-1} q \#_n)$ and $(q,a) \rightarrow a$, for
  every $a \in \Sigma$.  Finally, $(d_{\text{out}},s)$ defines a tree
  of depth two as follows:
  \begin{multline*}
    d_{\text{out}}(s)=\{\#_0w_1\#_1w_2\#_2\cdots\#_{n-1}w_n\#_n\mid {}\\
    \exists j\in\{1,\ldots,n\} \text{ such that $M_j$ does not accept
      $w_j$}\}.
  \end{multline*}
  Clearly, $d_{\text{out}}(s)$ can be represented by a DFA whose size
  is polynomial in the sizes of the $M_i$'s.  Indeed, the DFA just
  simulates every $M_i$ on the string following $\#_{i-1}$, until it
  encounters $\#_i$. It then verifies that at least one $M_i$ rejects.
  
  It is easy to see that this reduction can be carried out by a
  deterministic logspace algorithm.

  (2) This is an easy reduction from the universality problem of an
  NFA $N$ with alphabet $\{0,1\}$. The latter problem is \pspace-hard,
  as shown in Corollary~\ref{lem:tool-fix-alph}(2) in the Appendix.
  Again, the input DTD $(d_\text{in},s)$ defines a tree of depth two
  where $d_\text{in}(s) = (0+1)^*$. The tree transducer is the
  identity transformation. The output DTD $d_\out$ has as start symbol
  $s$ and $d_\out(s) = L(N)$.  Hence, this instance typechecks if and
  only if $\{0,1\}^* \subseteq L(N)$.

  This reduction can be carried out by a deterministic logspace
  algorithm.  \hfill $\Box$
\end{proof}

\subsection{Non-deleting: Fixed Output Schema}

Again, upper bounds carry over
from~\cite{martensneventcs05}.  Also, when the
output DTD is a DTD(NFA), we can convert it into an equivalent
DTD(DFA) in constant time. As the \ptime typechecking algorithm for
TC[$\cT_{nd,bc}$,DTD(DFA)] in~\cite{martensneventcs05} also works when
the input DTD is a DTD(NFA), we have that the problem
TC$^o$[$\cT_{nd,bc}$,DTD(NFA)] is in \ptime. As the \ptime-hardness
proof for TC[$\cT_{nd,bc}$,DTD(DFA)] in~\cite{martensneventcs05} uses
a fixed output schema, we immediately obtain the following.

\begin{theorem}
  $TC^o[\cT_{nc,bc},DTD(NFA)] $ is \ptime-complete.
\end{theorem}

The lower bound in the presence of tree automata will be discussed in
Section~\ref{sec:fixed:io}. The case requiring some real work is
$TC^o[\cT_{nd,uc}$, DTD(DFA)$]$.

\begin{theorem}
$TC^o[\cT_{nd,uc}$, DTD(DFA)$]$ is \pspace-complete.
\end{theorem}
\begin{proof}
  In~\cite{martensneventcs05}, it was shown that the problem is in
  \pspace. We proceed by showing \pspace-hardness.

  We use a \logspace reduction from the corridor tiling
  problem~\cite{Chlebus86}. Let $(T,V, \allowbreak H, \allowbreak \bar
  \vartheta, \bar \beta)$ be a tiling system, where $T =
  \{\vartheta_1,\ldots,\vartheta_k\}$ is the set of tiles, $V
  \subseteq T \times T$ and $H \subseteq T \times T$ are the sets of vertical
  and horizontal constraints respectively, and $\bar \vartheta$ and
  $\bar \beta$ are the top and bottom row, respectively. Let $n$ be
  the width of $\bar \vartheta$ and $\bar \beta$. The tiling system
  has a solution if there is an $m\in\nat$ such that the space
  $m\times n$ ($m$ rows and $n$ columns) can be correctly tiled with
  the additional requirement that the bottom and top row are $\bar
  \beta$ and $\bar \vartheta$, respectively.

  We define the input DTD $d_\text{in}$ over the alphabet
  $\Sigma:=\{(i,\vartheta_j)\mid j\in\{1,\ldots,k\},
  i\in\{1,\ldots,n\}\}\cup \{r\}$; $r$ is the start symbol. Define
  $$d_\text{in}(r)= \#\bar \beta \#\Bigl(
  \Sigma_1\cdot\Sigma_2\cdots\Sigma_n\#\Bigr)^*\bar \vartheta \#,$$
  where we
  denote by $\Sigma_{i}$ the set $\{(i,\vartheta_j)\mid
  j\in\{1,\ldots,k\}\}$.  Here, $\#$ functions as a row separator. For
  all other alphabet symbols $a \in \Sigma$,
  $d_\text{in}(a)=\varepsilon$. So, $d_{\text{in}}$ encodes all
  possible tilings that start and end with the bottom row $\bar \beta$ and
  the top row $\bar \vartheta$, respectively.

  We now construct a tree transducer $B = (Q_B,\Sigma,q^0_B,R_B)$ and
  an output DTD $d_\out$ such that $T$ has no correct corridor tiling
  if and only if $B$ typechecks with respect to $d_\text{in}$ and
  $d_\out$. Intuitively, the transducer and the output DTD have to
  work together to determine errors in input tilings. There can only
  be two types of error: two tiles do not match horizontally or two
  tiles do not match vertically. The main difficulty is that the
  output DTD is fixed and can, therefore, \emph{not} depend on the
  tiling system. The transducer is constructed in such a way that it
  prepares in parallel the verification for all horizontal and
  vertical constraints by the output schema. In particular, the
  transducer outputs specific symbols from a fixed set independent of
  the tiling system allowing the output schema to determine whether an
  error occurred.

  The state set $Q_B$ is partitioned into two sets, $Q_\text{hor}$ and
  $Q_\text{ver}$:
  \begin{itemize}
  \item $Q_\text{hor}$ is for the horizontal constraints: for every
    $i\in\{1,\ldots,n-1\}$ and $\vartheta\in T$, $q_{i,\vartheta} \in
    Q_\text{hor}$ transforms the rows in the tiling such that it is
    possible to check that when position $i$ carries a $\vartheta$,
    position $i+1$ carries a $\vartheta'$ such that
    $(\vartheta,\vartheta')\in H$; and,
  
  \item $Q_\text{ver}$ is for the vertical constraints: for every
    $i\in\{1,\ldots,n\}$ and $\vartheta\in T$, $p_{i,\vartheta} \in
    Q_\text{ver}$ transforms the rows in the tiling such that it is possible to
    check that when position $i$ carries a $\vartheta$, the next row
    carries a $\vartheta'$ on position $i$ such that
    $(\vartheta,\vartheta')\in V$.
  \end{itemize}

  The tree transducer $B$ always starts its transformation with the
  rule
  $$(q^0_B,r)\to r(w),$$
  where $w$ is the concatenation of all of the
  above states, separated by the delimiter $\$$. The other rules are
  of the following form:
  \begin{itemize}
  \item Horizontal constraints: for all $(j,\vartheta)\in \Sigma$ add
    the rule $(q_{i,\vartheta},(j,\vartheta'))\to \alpha$ where
    $q_{i,\vartheta} \in Q_\text{hor}$ and 
    $$\alpha = \left\{
      \begin{array}{ll}
        \texttt{trigger} & \text{if $j=i$ and
          $\vartheta=\vartheta'$}\\
        \texttt{other} & \text{if $j=i$ and $\vartheta\neq \vartheta'$}\\
        \texttt{ok} & \text{if $j=i+1$ and $(\vartheta,\vartheta') \in H$}\\
        \texttt{error} & \text{if $j=i+1$ and $(\vartheta,\vartheta')\not\in H$}\\
        \texttt{other} & \text{if $j\neq i$ and $j\neq i+1$}
      \end{array}
    \right.$$
    Finally, $(q_{i,\vartheta},\#)\to \text{\tt hor}$. 

    The intuition is as follows: if the $i$-th position in a row is
    labeled with $\vartheta$, then this position is transformed into
    \texttt{trigger}.  Position $i+1$ is transformed to \texttt{ok}
    when it carries a tile that matches $\vartheta$ horizontally.
    Otherwise, it is transformed to \texttt{error}. All other symbols
    are transformed into an \texttt{other}.  

    On a row, delimited by two \texttt{hor}-symbols, the output DFA
    rejects if and only if there is a \texttt{trigger} immediately
    followed by an \texttt{error}. When there is no \texttt{trigger},
    then position $i$ was not labeled with $\vartheta$.  So, the label
    \texttt{trigger} acts as a trigger for the output automaton.

  \item Vertical constraints: for all $(j,\vartheta)\in \Sigma$, add
    the rule $(p_{i,\vartheta},(j,\vartheta'))\to \alpha$ where
    $p_{i,\vartheta} \in Q_\text{ver}$ and $$
    \alpha = \left\{
      \begin{array}{ll}
        \texttt{trigger1} & \text{if $(j,\vartheta')=(i,\vartheta)$ and $(\vartheta,\vartheta) \in V$}\\
        \texttt{trigger2} & \text{if $(j,\vartheta')=(i,\vartheta)$ and $(\vartheta,\vartheta) \not\in V$}\\
        \texttt{ok} & \text{if $j=i$, $\vartheta\neq \vartheta'$, and $(\vartheta,\vartheta')\in V$}\\
        \texttt{error} & \text{if $j=i$, $\vartheta\neq \vartheta'$, and $(\vartheta,\vartheta')\not\in V$}\\
        \texttt{other} & \text{if $j\neq i$}
      \end{array}
    \right.$$
    Finally, $(p_{i,\vartheta},\#)\to \text{\tt ver}$.

    The intuition is as follows: if the $i$-th position in a row is
    labeled with $\vartheta$, then this position is transformed into
    \texttt{trigger1} when $(\vartheta,\vartheta) \in V$ and to
    \texttt{trigger2} when $(\vartheta,\vartheta) \not\in V$.  Here,
    both \texttt{trigger1} and \texttt{trigger2} act as a trigger for
    the output automaton: they mean that position $i$ was labeled with
    $\vartheta$. But no \texttt{trigger1} and \texttt{trigger2} can
    occur in the same transformed row as either $(\theta,\theta) \in
    V$ or $(\theta,\theta) \not \in V$. When position $i$ is labeled
    with $\vartheta'\neq \vartheta$, then we transform this position
    into \texttt{ok} when $(\vartheta,\vartheta')\in V$, and in
    \texttt{error} when $(\vartheta,\vartheta')\not\in V$. All other
    positions are transformed into \texttt{other}.
    
    The output DFA then works as follows. If a position is labeled
    \texttt{trigger1} then it rejects if there is a \texttt{trigger2}
    or a \texttt{error} occurring after the next \texttt{ver}.  If a
    position is labeled \texttt{trigger2}, then it rejects if there is
    a \texttt{trigger2} or a \texttt{error} occurring after the next
    \texttt{ver}. Otherwise, it accepts that row.
  \end{itemize}

  By making use of the delimiters \texttt{ver} and \texttt{hor}, both
  above described automata can be combined into one automaton, taking
  care of the vertical and the horizontal constraints. This automaton
  resets to its initial state whenever it reads the delimiter symbol
  $\$$.  Note that the output automaton is defined over the fixed
  alphabet $\{\texttt{trigger}, \texttt{trigger1}, \texttt{trigger2},
  \texttt{error}, \texttt{ok}, \texttt{other}, \texttt{hor},
  \texttt{ver},\$\}$. \hfill $\Box$
\end{proof}

Although the results in~\cite{martensneventcs05} were formulated in
the context of variable schemas, the proofs for bounded copying,
non-deleting tree transducers with DTD(SL) and with DTD(DFA) schemas
actually used a fixed output schema. We can therefore sharpen these
results as follows.
\begin{theorem}
  \begin{enumerate}
  \item $TC^o[\cT_{nd,bc}$, DTD(SL)$]$ is \conp-complete;
  \item $TC^o[\cT_{nd,bc}$, DTD(DFA)$]$ is \ptime-complete.
  \end{enumerate}
\end{theorem}

\subsection{Non-deleting: Fixed Input and Output Schema}
\label{sec:fixed:io}

We turn to the case where both input and output schemas are fixed. The
following two theorems give us several new tractable cases.

\begin{theorem}\label{theo:fix-io-nd-bc-nl-hard}
  \begin{enumerate}
  \item $TC^{io}[\cT_{nd,bc}$, DTD(SL)$]$ is \nlogspace-complete.
  \item $TC^{io}[\cT_{nd,bc}$, DTD(DFA)$]$ is \nlogspace-complete.
  \end{enumerate}
\end{theorem}
\begin{proof}
  For both problems, membership in \nlogspace follows from
  Theorem~\ref{theo:fix-io-nd-c-fa}. Indeed, every DTD(SL) can be
  rewritten into an equivalent DTD(NFA) in constant time as the input
  and output schemas are fixed.

  We proceed by showing \nlogspace-hardness. We say that an NFA $N =
  (Q_N, \Sigma, \delta_N, I_N, F_N)$ has \emph{degree of
    nondeterminism 2} if \emph{(i)} $I_N$ has at most two elements and
  \emph{(ii)} for every $q \in Q_N$ and $a \in \Sigma$, the set
  $\delta_N(q,a)$ has at most two elements.  We give a \logspace
  reduction from the emptiness problem of an NFA with alphabet
  $\{0,1\}$ and a degree of nondeterminism 2 to the typechecking
  problem.  According to Lemma~\ref{lem:NFA-01-nd2-nlog} in the
  Appendix, this problem is \nlogspace-hard.  Intuitively, the input
  DTD will define all possible strings over alphabet $\{0,1\}$.  The
  tree transducer simulates the NFA and outputs ``accept'' if a
  computation branch accepts, and ``error'' if a computation branch
  rejects. The output DTD defines trees where all leaves are labeled
  with ``error''.

  More concretely, let $N = (Q_N, \{0,1\}, \delta_N, \{q^0_N\}, F_N)$
  be an NFA with degree of nondeterminism 2. The input DTD
  $(d_\text{in},r)$ defines all unary trees, where the unique leaf is
  labeled with a special marker $\#$.  That is, $d_\text{in}(r) =
  d_\text{in}(0) = d_\text{in}(1) = (0+1+\#)$ and $d_\text{in}(\#) =
  \varepsilon$. Note that these languages can be defined by
  SL-formulas or DFAs which are sufficiently small for our purpose.

  Given a tree $t = r(a_1(\cdots(a_n(\#))\cdots))$, the tree
  transducer will simulate every computation of $N$ on the string
  $a_1\cdots a_n$.  The tree transducer $T = (Q_T,\{r, \#, 0,1,
  \text{error}, \text{accept}\}, q^0_T, R_T)$ simulates $N$'s
  nondeterminism by copying the remainder of the input twice in every
  step. Formally, $Q_T$ is the union of $\{q^0_T\}$ and $Q_N$, and
  $R_T$ contains the following rules:

  \begin{itemize}
  \item $(q^0_T,r) \to r(q^0_N)$. This rule puts $r$ as the root symbol of the
    output tree and starts the simulation of $N$.
  \item $(q_N,a) \to a(q^1_N, q^2_N)$, where $q_N \in Q_N$, $a \in
    \{0,1\}$ and $\delta_N(q_N,a) = \{q^1_N, q^2_N\}$. This rule does
    the actual simulation of $N$. By continuing in both states $q^1_N$
    and $q^2_N$, we simulate \emph{all} possible computations of $N$.
  \item $(q_N,a) \to \text{error}$ if $\delta_N(q_N,a) = \emptyset$.
    If $N$ rejects, we output the symbol ``error''.
  \item $(q_N,\#) \to \text{error}$ for $q_N \not \in F_N$; and
  \item $(q_N,\#) \to \text{accept}$ for $q_N \in F_N$. These last two
    rules verify whether $N$ is in an accepting state after reading
    the entire input string.
  \end{itemize}

  Notice that $T$ outputs the symbol ``error'' (respectively
  ``accept'') if and only if a computation branch of $N$ rejects
  (respectively accepts).

  The output of $T$ is always a tree in which only the symbols
  ``error'' and ``accept'' occur at the leaves. The output DTD then
  needs to verify that only the symbol ``error'' occurs at the leaves.
  Formally, $d_\out(r) = d_\out(0) = d_\out(1) =
  \{0,1,\text{error}\}^+$ and $d_\out(\text{error}) = \varepsilon$.
  Again, these languages can be defined by sufficiently small
  SL-formulas or DFAs.

  It is easy to see that the reduction only requires logarithmic
  space.  \hfill $\Box$
\end{proof}

\begin{theorem}\label{theo:fix-io-nd-c-fa}
  $TC^{io}[\cT_{nd,uc}$, DTD(NFA)$]$ is \nlogspace-complete.
\end{theorem}
\begin{proof}
  The \nlogspace-hardness of the problem follows from
  Theorem~\ref{theo:fix-io-nd-bc-nl-hard}(b), where it shown that the
  problem is already \nlogspace-hard when DTD(DFA)s are used as input
  and output schema.

  We show that the problem is also in \nlogspace. Thereto, let $T =
  (Q_T, \Sigma, q^0_T,\allowbreak R_T)$ be the tree transducer, and
  let $(d_\text{in},r)$ and $d_\out$ be the input and output DTDs,
  respectively. As both $d_\text{in}$ and $d_\out$ are fixed, we can
  assume without loss of generality that they are reduced.\footnote{In
    general, reducing a DTD(NFA) is \ptime-complete
    (Section~\ref{sec:DTD-TA}).} For the same reason, we can also
  assume that the NFAs in $d_\text{in}$ and $d_\out$ are determinized.
 
  We guess a sequence of state-label pairs $(p_0,a_0),(p_1,a_1),
  \ldots, (p_n,a_n)$ where $n < |Q_T||\Sigma|$ such that
  \begin{itemize}
  \item $(p_0,a_0) = (q^0_T,r)$; and
  \item for every pair $(p_i,a_i)$, $p_{i+1}$ occurs in rhs$(p_i,a_i)$
    and $a_{i+1}$ occurs in some string in $L(d_\text{in}(a_i))$. 
  \end{itemize}
  Each time we guess a new pair in this sequence, we forget the
  previous one, so that we only keep a state, an alphabet symbol, a
  counter, and the binary representation of $|Q_T||\Sigma|$ on tape.
  
  For simplicity, we write $(p_n,a_n)$ as $(p,a)$ in the remainder of
  the proof. We guess a node $u \in \Dom(\rhs(p,a))$. Let $b =
  \lab^{\rhs(p,a)}(u)$ and let $z_0 q_1 z_1 \cdots q_k z_k$ be the
  concatenation of $u$'s children, where every $z_0, \ldots, z_k \in
  \Sigma^*$ and every $q_1,\ldots,q_k \in Q_T$, then we want to check
  whether there exists a string $w \in d_\text{in}(a)$ such that $z_0
  q_1[w] z_1 \cdots q_k[w] z_k$ is not accepted by $d_\out(b)$. Recall
  from Section~\ref{sec:transducers} that, for a state $q \in Q_T$, we
  denote by $q[w]$ the homomorphic extension of $q[c]$ for $c \in
  \Sigma$, which is $\text{top}(\text{rhs}(q,c)))$ in the case of
  non-deleting tree transducers. We could do this by guessing $w$ one
  symbol at a time and simulating $k$ copies of $d_\out(b)$ and one
  copy of $d_\text{in}(a)$ in parallel, like in the proof of
  Theorem~\ref{theo:fix-i-nd-bc-dfa}. However, as $k$ is not fixed,
  the algorithm would use superlogarithmic space.

  So, we need a different approach. To this end, let $A =
  (Q_\text{in},\Sigma,\delta_{\text{in}},q^0_\text{in},F_\text{in})$
  and
  $B=(Q_\text{out},\Sigma,\delta_{\text{out}},q^0_\text{out},F_\text{out})$
  be the DFAs accepting $d_\text{in}(a)$ and $d_\text{out}(b)$,
  respectively. To every $q \in Q_T$, we associate a function $$f_q :
  Q_\out \times \Sigma \to Q_\out : (p',c) \mapsto \hat
  \delta_\text{out}(p',q[c]),$$
  where $\hat \delta_\text{out}$ denotes
  the canonical extension of $\delta_\text{out}$ to strings in
  $\Sigma^*$. Note that there are maximally
  $|Q_\out|^{|Q_\out||\Sigma|}$ such functions.  Let $K$ be the
  cardinality of the set $\{f_q \mid q \in Q_T\}$. Hence, $K$ is
  bounded from above by $|Q_\out|^{|Q_\out||\Sigma|}$, which is a
  constant (with respect to the input). Let $f_1,\ldots,f_K$ an
  arbitrary enumeration of $\{f_q \mid q \in Q_T\}$.

  The typechecking algorithm continues as follows. We start by writing
  the $(1+K\cdot|Q_\out|)$-tuple
  $(q^0_\text{in},q'_1,\ldots,q'_{|Q_\out|},\ldots,q'_1,\ldots,q'_{|Q_\out|})$
  on tape, where $Q_\out = \{q'_1,\ldots,q'_{|Q_\out|}\}$. We will
  refer to this tuple as the tuple $\bar{p} := (p'_0,\ldots,p'_{K\cdot|Q_\out|})$.
  We explain how we update $\bar{p}$ when guessing $w$ symbol by
  symbol.  Every time when we guess the next symbol $c$ of $w$, we
  overwrite the tuple $\bar{p}$ by
  \begin{multline*}
    \big(\delta_\text{in}(p'_0,c), f_1(p'_1,c), \ldots,
    f_1(p'_{|Q_\out|},c), \ldots \\ \ldots, f_K(p'_{(K-1)\cdot|Q_\out|+1},c),
    \ldots, f_K(p'_{K\cdot|Q_\out|},c)\big).
  \end{multline*}
  Notice that there are at most $|Q_\text{in}| \cdot K \cdot
  |Q_\out|^2$ different $(K\cdot|Q_\out|+1)$-tuples of this form. We
  nondeterministically determine when we stop guessing symbols of $w$.

  It now remains to verify whether $w$ was indeed a string such that
  $w \in d_\text{in}(a)$ and $z_0 q_1[w] z_1 \cdots q_k[w] z_k \not
  \in d_\out(b)$. The former condition is easy to test: we simply have
  to test whether $p'_0 \in F_\text{in}$. To test the latter
  condition, we read the string $z_0 q_1 z_1 \cdots q_k z_k$ from left
  to right while performing the following tests. We keep a state of
  $d_\out(b)$ in memory and refer to it as the ``current state''.

  \begin{enumerate}
  \item The initial current state is $q_\out^0$.
  \item If the current state is $p'$ and we read $z_j$, then we
    change the current state to $\hat \delta_\text{out}(p',z_j)$.
  \item If the current state is $p'$ and we read $q_j$, then we change
    the current state to $p'_i$ in $\bar{p}$, where for $i$, the
    following condition holds.  Let $\ell,m = 1,\ldots,K\cdot|Q_\out|$
    be the smallest integers such that
    \begin{itemize}
    \item $p' = q'_\ell$ in $Q_\text{out}$, and
    \item $f_{q_j} = f_m$. 
    \end{itemize}
    Then $i = (m-1)K + \ell$. 

    Note that deciding whether $p' = q'_\ell$ and $f_{q_j} = f_m$ can
    be done deterministically in logarithmic space, as the output
    schema is fixed.  Consequently, $i$ can also be computed in
    constant time and space.
  \item We stop and accept if the current state is a non-accepting
    state after reading $z_k$.
  \end{enumerate}
  \hfill $\Box$
\end{proof}

\begin{theorem}\label{lem:fix-io-nd-bc-TA}
  $TC^{io}[\cT_{nd,bc}$, DTA(DFA)$]$ is \exptime-complete.
\end{theorem}
\begin{proof}
  The proof is quite analogous to the proof of
  Theorem~\ref{theo:fix-io-dw2-bc-exptime}. As deletion is now
  disallowed, whereas it was allowed in
  Theorem~\ref{theo:fix-io-dw2-bc-exptime}, we need to define the
  rules of the transducer $T = (Q_T, \Sigma_T,
  q^\varepsilon_\text{copy}, R_T)$ differently.

  The language defined by the input schema is exaclty the same as in
  Theorem~\ref{theo:fix-io-dw2-bc-exptime}.  The transition rules in
  $R_T$ are defined as follows:
  \begin{itemize}
  \item $(q^\varepsilon_\text{copy},s) \to s(q^0_\text{copy}
    q^1_\text{copy})$;
  \item $(q^i_\text{copy},\#) \to \#(q^{i0}_\text{copy}
    q^{i1}_\text{copy})$ for $i \in D(\lceil \log n \rceil - 1) -
    \{\varepsilon\}$; 
  \item $(q^i_\text{copy},\#) \to \#(\text{start}^\ell_k)$, where $i \in
    D(\lceil \log n \rceil) - D(\lceil \log n \rceil -1)$, and $i$ is
    the binary representation of $k$;
  \item $(q^i_\text{copy},a) \to \text{error}$ for $a \in \Sigma$ and
    $i \in D(\lceil \log n \rceil)$;
  \item $(\text{start}^\ell_k,\#) \to \text{error}$ for all $k =
    1,\ldots,2^{\lceil \log n \rceil}$;
  \item $(q^\ell,a_r) \to \varepsilon$ and $(q^r,a_\ell) \to
    \varepsilon$ for all $q \in Q_j$, $j = 1,\ldots,n$;
  \item $(q^\ell,a_\ell) \to a_\ell(q_1^\ell q_2^r)$ and $(q^r,a_r) \to
    a_r(q_1^\ell q_2^r)$, for every $q \in Q_i$, $i = 1,\ldots,n$, such
    that $\delta_i(q,a) = q_1q_2$, and $a$ is an internal symbol;
  \item $(q^\ell,a_\ell) \to a_\ell(q_1^\ell)$ and $(q^r,a_r) \to a_r(q_1^\ell)$,
    for every $q \in Q_i$, $i = 1,\ldots,n$, such that $\delta_i(q,a)
    = q_1$ and $a$ is an internal symbol;
  \item $(q^\ell,a_\ell) \to \varepsilon$ and $(q^r,a_r) \to
    \varepsilon$ for every $q \in Q_i$, $i = 1,\ldots,n$, such that
    $\delta_i(q,a) = \varepsilon$ and $a$ is a leaf symbol; and
  \item $(q^\ell,a_\ell) \to $ error and $(q^r,a_r) \to$ error for
    every $q \in Q_i$, $i = 1,\ldots,n$, such that $\delta_i(q,a)$ is
    undefined.
  \end{itemize}
  It is straightforward to verify that, on input
  $s(\#(\#(\cdots\#(t))))$, $T$ performs the identity transformation
  if and only if there are $\lceil \log n \rceil$ $\#$-symbols in the
  input and $t \in L(A_1) \cap \cdots \cap L(A_n)$. All other
  outputs contain at least one leaf labeled ``error''.

  Finally, the output tree automaton accepts all trees with at least
  one leaf that is labeled ``error''. So the only counterexamples for
  typechecking are those trees that are accepted by all automata.
  
  It is easy to see that the reduction can be carried out in
  deterministic logarithmic space, that $T$ has copying width 2, and
  that the input and output schemas do not depend on $A_1,\ldots,A_n$.
  \hfill $\Box$
\end{proof}
\section{Conclusion} \label{sec:conclusion} We considered the
complexity of typechecking in the presence of fixed input and/or
output schemas.  We have settled an open question
in~\cite{martensneventcs05}, namely that TC$[\cT_{nd,bc},DTA]$ is
\exptime-complete. 

In comparison with the results in~\cite{martensneventcs05}, fixing
input and/or output schemas only lowers the complexity in the presence
of DTDs and when deletion is disallowed. Here, we see that
the complexity is lowered when
\begin{enumerate}
\item the input schema is fixed, in the case of DTD(SL)s;
\item the input schema is fixed, in the case of DTD(DFA)s;
\item the output schema is fixed, in the case of DTD(NFA)s; and 
\item both input and output schema are fixed, in all cases.
\end{enumerate}
In all of these cases, the complexity of the typechecking problem is
in polynomial time.

It is striking, however, that in many cases, the complexity of
typechecking does not decrease significantly by fixing the input
and/or output schema, and most cases remain intractable. We have to
leave the precise complexity (that is, the \ptime-hardness) of
$TC^i[\cT_{nd,uc}$, DTD(SL)$]$ as an open problem.

\section*{Acknowledgments}
We thank Giorgio Ghelli for raising the question about the complexity
of typechecking in the setting of a fixed output schema. We also thank
Joos Heintz for providing us with a useful reference to facilitate the
proof of Proposition~\ref{prop:extended-integer-programming}.

\appendix
\section*{Appendix: Definitions and Basic Results}

\newcommand{\trenc}{\text{tree-enc}}

\label{sec:app:basic}

The purpose of this Appendix is to prove some lemmas that we use in
the body of the paper.  We first introduce some notations and
definitions needed for the propositions and proofs further on in this
Appendix. We also survey some complexity bounds on decisions problems
concerning automata that are used throughout the paper.

We show that the complexities of the classical decision problems of
string and tree automata are preserved when the automata operate over
fixed alphabets.  We will consider the following decision problems for
string automata:
\begin{description}
\item[Emptiness:] Given an automaton $A$, is $L(A) = \emptyset$?
\item[Universality:] Given an automaton $A$, is $L(A) = \Sigma^*$?
\item[Intersection emptiness:] Given the automata $A_1, \ldots, A_n$,
  is $L(A_1) \cap \cdots \cap L(A_n) = \emptyset$?
\end{description}
The corresponding decision problems for tree automata are defined
analogously.  

We associate to each label $a \in\Sigma$ a unique binary string
$\enc(a) \in \{0,1\}^*$ of length $\lceil \log |\Sigma| \rceil$. For a
string $s=a_1\cdots a_n$, $\enc(s)=\enc(a_1)\cdots \enc(a_n)$. This
encoding can be extended to string languages in the obvious way.

We show how to extend the encoding ``enc'' to trees over alphabet
$\{0,1,0',1'\}$.  Here, 0 and 1 are internal labels, while $0'$ and
$1'$ are leaf labels.  Let $\enc(a)=b_1\cdots b_k$ for $a\in\Sigma$.
Then we denote by $\trenc(a)$ the unary tree $b_1(b_2(\cdots (b_k))$,
if $a$ is an internal label, and the unary tree $b_1(b_2(\cdots
(b'_k))$, otherwise. Then, the $\enc$-fuction can be extended to trees
as follows: for $t=a(t_1\cdots t_n)$,
$$\enc(t)=\trenc(a)(\enc(t_1)\cdots \enc(t_n)).$$
Note that we abuse
notation here. The hedge $\enc(t_1)\cdots \enc(t_n)$ is intended to be
the child of the leaf in $\trenc(a)$.  The encoding can be extended to
tree languages in the obvious way.

\begin{proposition}\label{lem:arb->fixed alphabet}
  Let $B$ be a TDBTA. Then there is a TDBTA $B'$ over the alphabet
  $\{0,1,0',1'\}$ such that $L(B')=\enc(L(B))$. Moreover, $B'$ can be
  constructed from $B$ in \logspace.
\end{proposition}
\begin{proof}
  Let $B=(Q_B, \Sigma_B, \delta_B, F_B)$ be a TDBTA.  Let $k := \lceil
  \log |\Sigma_B| \rceil$. We define $B' = (Q_{B'}, \{0,1,0',1'\},
  \allowbreak \delta_{B'}, F_{B'})$. Set $Q_{B'}=\{q_x \mid q \in Q_B$
  and $x$ is a prefix of $\text{enc}(a)$, where $a \in \Sigma_B\}$ and
  $F_{B'}=\{q_\varepsilon \mid q \in F_B\}$. To define the transition
  function, we introduce some notation. For each $a \in \Sigma$ and
  $i, j = 1,\ldots,\lceil \log |\Sigma_B| \rceil$, denote by
  $a[i$\,:\,$j]$ the substring of $\text{enc}(a)$ from position $i$ to
  position $j$ (we abbreviate $a[i$\,:\,$i]$ by $a[i]$).  For each
  transition $\delta_B(q,a) = q^1q^2$, add the transitions
  $\delta_{B'}(q_\varepsilon,a[1]) = q_{a[1]}$,
  $\delta_{B'}(q_{a[1]},a[2]) = q_{a[1:2]}, \ldots,
  \delta_{B'}(q_{a[1:k-1]},a[k]) = q^1_\varepsilon q^2_\varepsilon$.
  Other transitions are defined analogously.  Clearly, $B'$ is a
  TDBTA, $L(B')=\enc(L(B))$, and $B'$ can be constructed from $B$ in
  \logspace. \hfill $\Box$
\end{proof}

It is straightforward to show that Proposition~\ref{lem:arb->fixed
  alphabet} also holds for NFAs and DFAs (the proofs are analogous).
It is immediate from Proposition~\ref{lem:arb->fixed alphabet}, that
lower bounds of decision problems for automata over arbitrary
alphabets~\cite{kozen77,seidl94,StockmeyerM73} carry over to automata
working over fixed alphabets. Hence, we obtain the following corollary
to Proposition~\ref{lem:arb->fixed alphabet}:

\begin{corollary} \label{lem:tool-fix-alph} Over the alphabet
  $\{0,1\}$, the following statements hold:
  \begin{enumerate}
  \item Intersection emptiness of an arbitrary number of DFAs is
    \pspace-hard.
  \item Universality of NFAs  is \pspace-hard.
  \end{enumerate}
  Over the alphabet $\{0,1,0',1'\}$, the following statement holds:
  \begin{enumerate}
  \item[(3)] Intersection emptiness of an arbitrary number of TDBTAs is
    \exptime-hard.
  \end{enumerate}
\end{corollary}

Lemma~\ref{lem:NFA-01-nd2-nlog} now immediately follows from
\nlogspace-hardness of the reachability problem on graphs with
out-degree 2~\cite{johnson_handbooktcs}.
\begin{lemma}\label{lem:NFA-01-nd2-nlog}
  The emptiness problem for an NFA with alphabet $\{0,1\}$ degree of
  nondeterminism 2 is \nlogspace-hard.
\end{lemma}

We now aim at proving
Proposition~\ref{prop:extended-integer-programming}, which states that
we can find integer solutions to arbitrary Boolean combinations of
linear (in)equalities in polynomial time, when the number of variables
is fixed.  To this end, we revisit a lemma that is due to Ferrante and
Rackoff~\cite{ferranterackoff}.

First, we need some definitions. We define logical formulas with
variables $x_1, x_2, \ldots $ and linear equations with factors in
$\rat$. A \emph{term} is an expression of the form $a_1/b_1$, $a_1/b_1
x_1 + \cdots + a_n/b_n x_n$, or $a_1/b_1 x_1 + \cdots +
a_{n-1}/b_{n-1} x_{n-1} + a_{n}/b_{n}$ where $a_i, b_i \in \nat$ for
$i = 1,\ldots,n$.  An \emph{atomic formula} is either the string
``true'', the string ``false'', or a formula of the form $\vartheta_1
= \vartheta_2$, $\vartheta_1 < \vartheta_2$, or $\vartheta_1 >
\vartheta_2$. A \emph{formula} is built up from atomic formulas using
conjunction, disjunction, negation, and the symbol $\exists$ in the
usual manner.  Formulas are interpreted in the obvious manner over
$\rat$. For instance, the formula $\lnot \exists x_1,x_2 \ (x_1 < x_2)
\land \lnot \big( \exists x_3\ (x_1 < x_3 \land x_3 < x_2) \big)$
states that for every two different rational numbers, there exists a
third rational number that lies strictly between them.

The \emph{size} of a formula $\Phi$ is the sum of the number of
brackets, Boolean connectives, the sizes of the variables, and the
sizes of all rational constants occurring in $\Phi$.  Here, we assume
that all rational constants are written as $a/b$, where $a$ and $b$
are integers, written in binary notation.  We assume that variables
are written as $x_i$, where $i$ is written in binary notation.

\begin{lemma}[Lemma 1 in~\cite{ferranterackoff}] \label{lem:ferranterackoff}
  Let $\Phi(x_1,\ldots,x_n)$ be a quantifier-free formula. Then there
  exists a \ptime procedure for obtaining another quantifier-free
  formula, $\Phi'(x_1,\ldots,x_{n-1})$, such that
  $$\Phi'(x_1,\ldots,x_{n-1}) \text{ is equivalent to } \exists x_n
  \Phi(x_1,\ldots,x_n).$$
\end{lemma}
\begin{proof}
  Let $\Phi(x_1,\ldots,x_n)$ be a quantifier-free formula.
  \begin{description}
  \item[Step 1:] \emph{Solve for $x_n$} in each atomic formula of
    $\Phi$. That is, obtain a quantifier-free formula,
    $\Psi^n(x_1,\ldots,x_n)$, such that every atomic formula of
    $\Psi^n$ either does not involve $x_n$ or is of the form (i) $x_n
    < \vartheta$, (ii) $x_n > \vartheta$, or (iii) $x_n = \vartheta$,
    where $\vartheta$ is a term not involving
    $x_n$. 
  \item[Step 2:] We now make the following definitions:\\ Given
    $\Psi^n(x_1,\ldots,x_n)$, to get $\Psi_{-\infty}^n(x_1,\ldots,x_{n-1})$,
    respectively, $\Psi_{\infty}^n(x_1,\ldots,x_{n-1})$, replace
    \begin{center}
      \begin{tabular}{l}
        $x_n < \vartheta^n$ in $\Psi$ by ``true'' (respectively, ``false'');\\
        $x_n > \vartheta^n$ in $\Psi$ by ``false'' (respectively, ``true''); and,\\
        $x_n = \vartheta^n$ in $\Psi$ by ``false'' (respectively, ``false'').
      \end{tabular}
    \end{center}
    The intuition is that, for any rational numbers $r_1, \ldots,
    r_{n-1}$, if $r$ is a sufficiently small rational number, then
    $\Psi^n(r_1,\ldots,r_{n-1},r)$ and
    $\Psi_{-\infty}^n(r_1,\ldots,r_{n-1})$ are equivalent. A similar
    statement can be made for $\Psi_{\infty}^n$ for $r$ sufficiently
    large.
  \item[Step 3:] We will now eliminate the quantifier from $\exists
    x_n \Psi(x_1, \ldots, x_n)$. Let $U$ be the set of all terms
    $\vartheta$ (not involving $x_n$) such that $x_n > \vartheta$,
    $x_n < \vartheta$, or $x_n = \vartheta$ is an atomic formula of
    $\Psi$.  Lemma 1.1 in~\cite{ferranterackoff} then shows that
    $\exists x_n \Psi(x_1, \ldots, x_n)$ is equivalent to the
    quantifier-free formula $\Phi'(x_1, \ldots, x_{n-1})$ defined to
    be $$\Psi_{-\infty}^n \lor \Psi_{\infty}^n \lor
    \bigvee_{\vartheta, \vartheta' \in U} \Psi^{n, \vartheta,
      \vartheta'},$$
    where $\Psi^{n,\vartheta, \vartheta'} = \Psi^n\big(
    x_1, \ldots, x_{n-1}, \frac{\vartheta + \vartheta'}{2} \big)$. 
  \hfill $\Box$
  \end{description} 
\end{proof}

The following proposition is implicit in the work by Ferrante and
Rackoff~\cite{ferranterackoff}, we prove it for completeness:

\begin{proposition} \label{prop:rational} 
  Let $\Phi(x_1,\ldots,x_n)$
  be a quantifier-free formula. If $n$ is fixed, then satisfiability
  of $\Phi$ over $\rat$ can be decided in \ptime.  Moreover, if $\Phi$
  is satisfiable, we can find $(v_1,\ldots,v_n) \in \rat^n$ such that
  $\Phi(v_1,\ldots,v_n)$ is true in polynomial time.
\end{proposition}
\begin{proof}
  We first show that satisfiability can be decided in \ptime. To this
  end, we simply iterate over the three steps in the proof of
  Lemma~\ref{lem:ferranterackoff} until we obtain a formula without
  variables. Hence, in each iteration, one variable $x_i$ is
  eliminated from $\Phi$.  For every $i = 1, \ldots, n$, let $\Phi^i$
  be the formula obtained after eliminating variable $x_i$.
  
  Notice that, in each iteration of the algorithm, the number of
  atomic formulas grows quadratically when going from $\Phi^i$ to
  $\Phi^{i-1}$. However, as there are only a constant number of
  iterations, the number of atomic formulas in the resulting formula
  $\Phi^1$ is still polynomial. Moreover, Ferrante and Rackoff show
  that the absolute value of every integer constant occurring in any
  rational constant in $\Phi^i$ is at most $(s_0)^{14^n}$, where $s_0$
  is the largest absolute value of any integer constant occurring in
  any rational constant in $\Phi$ (\mbox{cfr.} page 73
  in~\cite{ferranterackoff}).  As $n$ is a fixed number, we can decide
  whether $\Phi^1$ is satisfiable in polynomial time.
  
  Suppose that $\Phi$ is satisfiable.  We now show how we can
  construct $(v_1,\ldots,v_n) \in \rat^n$ in polynomial time such that
  $\Phi(v_1,\ldots,v_n)$ is true. For a term $\vartheta$ using
  variables $x_1,\ldots,x_i$, we denote by
  $\vartheta(v_1,\ldots,v_{i-1})$ the rational number obtained by
  replacing the variables $x_1,\ldots,x_i$ in $\vartheta$ by
  $v_1,\ldots,v_{i-1}$ and evaluating the resulting expression.

  For every $i = 1, \ldots,n$, we
  construct $v_i$ from $\Psi^i$, $\Psi^i_{-\infty}$,
  $\Psi^i_{\infty}$, and $\Psi^{i,\vartheta, \vartheta'}$ (which are
  defined in the proof of Lemma~\ref{lem:ferranterackoff}) as follows:
  \begin{enumerate}[(1)]
  \item If $\Psi^{i,\vartheta, \vartheta'}$ is satisfiable, then $v_i
    = \frac{\vartheta(v_1,\ldots,v_{i-1}) +
      \vartheta'(v_1,\ldots,v_{i-1})}{2}$.
  \item Otherwise, if $\Psi^i_{\infty}$ is satisfiable, then $v_i =
    \max\{\vartheta(v_1,\ldots,v_{i-1}) \mid x_i < \vartheta$ or $x_i
    > \vartheta$ or $ x_i = \vartheta$ is an atomic formula in
    $\Psi^i\} + 1$.
  \item Otherwise, if $\Psi^i_{-\infty}$ is satisfiable, then $v_i =
    \min\{\vartheta(v_1,\ldots,v_{i-1}) \mid x_i < \vartheta$ or $x_i
    > \vartheta$ or $ x_i = \vartheta$ is an atomic formula in
    $\Psi^i\} - 1$.
  \end{enumerate}
  
  It remains to show that we can represent every $v_i$ in a polynomial
  manner. 

  In the proof of Lemma 2 in~\cite{ferranterackoff}, Ferrante and
  Rackoff show that, if $w_i$ is the maximum absolute value of any
  integer occurring in the definition of $v_1,\ldots, v_i$, then we
  have the recurrence
  $$w_{i+1} \leq (s_0)^{2^{cn}} \cdot (w_i)^i,$$
  for a constant $c$
  and $s_0$ defined as before. Let $c' = 2^{cn}$.  Hence, the maximum
  number of bits needed to represent the largest integer in $v_{i+1}$
  is $\log \big((s_0)^{c'} \cdot (w_i)^i \big) = c' \log(s_0) \cdot i
  \log(w_i)$, which is polynomially larger than $\log(w_i)$, the
  number of bits needed to represent the largest integer in $v_i$.
  
  As we only have a constant number of iterations, the number of bits
  needed to represent the largest integer occurring in the definition
  of $v_n$ is also polynomial.  \hfill $\Box$
\end{proof}

The following proposition is a generalization of a well-known theorem
by Lenstra which states that there exists a polynomial time algorithm
to find an integer solution for a \emph{conjunction} of linear
(in)equalities with rational factors and a fixed number of
variables~\cite{lenstraIP}.
\begin{proposition} \label{prop:extended-integer-programming} There
  exists a \ptime algorithm that decides whether a Boolean combination
  of linear (in)equalities with rational factors and a fixed number of
  variables has an integer solution.
\end{proposition}
\begin{proof}
  Note that we cannot simply put the Boolean combination into
  disjunctive normal form, as this would lead to an exponential
  increase of its size.

  Let $\Phi(x_1,\ldots,x_n)$ be a Boolean combination of formulas
  $\varphi_1, \ldots, \varphi_m$ with variables $x_1,\ldots,x_n$ that
  range over $\integers$. Here, $n$ is a constant integer greater than
  zero. Without loss of generality, we can assume that every
  $\varphi_i$ is of the form $$k_{i,1} \times x_1 + \cdots + k_{i,n}
  \times x_n + k_i \geq 0,$$
  where $k_i, k_{i,1}, \ldots, k_{i,n} \in
  \rat$.

  We describe a \ptime procedure for finding a solution for
  $x_1,\ldots,x_n$, that is, for finding values $v_1,\ldots,v_n \in
  \integers$ such that $\Phi(v_1,\ldots,v_n)$ evaluates to true.

  First, we introduce some notation and terminology.  For every $i =
  1, \ldots, m$, we denote by $\varphi'_i$ the formula $k_{i,1} \times
  x_1 + \cdots + k_{i,n} \times x_n + k_i = 0$. In the following, we
  freely identify $\varphi'_i$ with the hyperplane it defines in
  $\reals^n$. For an $n$-tuple $\overline{y} = (y_1,\ldots,y_n) \in
  \rat^n$, we denote by $\varphi'_i(\overline{y})$ the rational number
  $k_{i,1} \times y_1 + \cdots + k_{i,n} \times y_n + k_i$.

  Given a set of hyperplanes $H$ in $\reals^n$, we say that $C
  \subseteq \reals^n$ is a \emph{cell} of $H$ when 
  \begin{enumerate}[(i)]
  \item for every hyperplane $\varphi'_i$ in $H$, and for every pair
    of points $\overline{y}$, $\overline{z} \in C$, we have that
    $\varphi'_i(\overline{y})\ \theta\ 0$ if and only if
    $\varphi'_i(\overline{z})\ \theta\ 0$, where, $\theta$ denotes
    ``$<$'', ``$>$'', or ``$=$''; and
  \item there exists no $C' \supsetneq C$ with property (i).
  \end{enumerate}

  Let $H$ be the set of hyperplanes $\{\varphi'_i \mid 1 \leq i \leq
  m\}$.

  We now describe the \ptime algorithm. The algorithm iterates over
  the following steps:
  \begin{enumerate}[(1)]
  \item Compute $(v'_1, \ldots, v'_n) \in \rat^n$ such that
    $\Phi(v'_1,\ldots,v'_n)$ is true.\footnote{Note that we abuse
      notation here, as the variables in $\Phi$ range over $\integers$
      and not $\rat$.} If no such $(v'_1, \ldots, v'_n)$ exists, the
    algorithm rejects.
  \item For every $\varphi'_i \in H$, let $\theta_i \in \{<,>,=\}$ be
    the relation such that $$k_{i,1} \times v'_1 + \cdots + k_{i,n}
    \times v'_n + k_i\ \theta_i \ 0.$$
    For every $i = 1, \ldots, m$,
    let $\varphi''_i = k_{i,1} \times x_1 + \cdots + k_{i,n} \times
    x_n + k_i\ \theta_i \ 0$. So, for every $i = 1, \ldots, m$,
    $\varphi''_i$ defines the half-space or hyperplane that contains
    the point $(v'_1,\ldots,v'_n)$.
    
    Let $\Phi'(x_1,\ldots,x_n)$ be the conjunction
    $$\bigwedge_{1 \leq i \leq n} \varphi''_i.$$
    Notice that the points
    satisfying $\Phi'(x_1,\ldots,x_n)$ are precisely the points in the
    cell $C$ of $H$ that contains $(v'_1,\ldots,v'_n)$.  
  \item Solve the \emph{integer programming problem} for
    $\Phi'(x_1,\ldots,x_n)$.  That is, find a $(v_1,\ldots,v_n) \in
    \integers^n$ such that $\Phi'(v_1,\ldots,v_n)$ evaluates to true.
  \item If $(v_1,\ldots,v_n) \in \integers^n$ exists, then write
    $(v_1,\ldots,v_n)$ to the output and accept.
  \item If $(v_1,\ldots,v_n) \in \integers^n$ does not exist, then
    overwrite $\Phi(x_1,\ldots,x_n)$ with $$\Phi''(x_1,\ldots,x_n) =
    \Phi(x_1,\ldots,x_n) \land \lnot \Phi'(x_1,\ldots,x_n)$$
    and go
    back to step (1).
  \end{enumerate}
  
  We show that the algorithm is correct.  Clearly, if the algorithm
  accepts, $\Phi$ has a solution. Conversely, suppose that $\Phi$ has
  a solution. Hence, the algorithm computes a value $(v'_1, \ldots,
  v'_n) \in \rat^n$ in step (1) of its first iteration.  It follows
  from the following two observations that the algorithm accepts:
  \begin{enumerate}[(i)]
  \item If the algorithm computes $(v'_1, \ldots, v'_n) \in \rat^n$ in
    step (1), and the cell $C$ of $H$ containing $(v'_1, \ldots,
    v'_n)$ also contains a point in $\integers^n$, then step (3) finds
    a solution $(v_1,\ldots,v_n) \in \integers^n$; and,
  \item If the algorithm computes $(v'_1, \ldots, v'_n) \in \rat^n$ in
    step (1), and the cell $C$ of $H$ containing $(v'_1, \ldots,
    v'_n)$ does \emph{not} contain a point in $\integers^n$, then step
    (3) does \emph{not} find a solution. By construction of $\Phi''$
    in step (5), the solutions to the formula $\Phi''$ are the
    solutions of $\Phi$, minus the points in $C$. As $C$ did not
    contain a solution, we have that $\Phi$ has a solution if and only
    if $\Phi''$ has a solution. Moreover, there exists no $(v''_1,
    \ldots, v''_n) \in C$ such that $\Phi''(v''_1,\ldots,v''_n)$
    evaluates to true.
  \end{enumerate}

  To show that the algorithm can be implemented to run in polynomial
  time, we first argue that there are at most a polynomial number of
  iterations. This follows from the observation in step (2) that the
  points satisfying $\Phi'(x_1,\ldots,x_n)$ are precisely all the
  points in a cell $C$ of $H$. Indeed, when we do not find a solution
  to the problem in step (3), we adapt $\Phi$ to exclude all the
  points in cell $C$ in step (5). Hence, in the following iteration,
  step (1) cannot find a solution in cell $C$ anymore. It follows that
  the number of iterations is bounded by the number of cells in $H$,
  which is $\Theta(m^n)$ (see, \mbox{e.g.}~\cite{buck}, or Theorem 1.3
  in~\cite{edelsbrunnerbook} for a more recent reference).
  
  Finally, we argue that every step of the algorithm can be computed
  in \ptime.
  
  Step (1) can be solved by the quantifier elimination method of
  Ferrante and Rackoff (Lemma~\ref{lem:ferranterackoff}).
  Proposition~\ref{prop:rational} states that we can find
  $(v'_1,\ldots,v'_n)$ in polynomial time.
  
  Step (2) is easily to be seen to be in \ptime: we only have to
  evaluate every $\varphi'_i$ once on $(v'_1,\ldots,v'_n)$.
  
  Step (3) can be executed in \ptime by Lenstra's algorithm for
  integer linear programming with a fixed number of
  variables~\cite{lenstraIP}.
  
  Step (4) is in \ptime (trivial).
  
  Step (5) replaces $\Phi(x_1,\ldots,x_n)$ by the formula
  $\Phi(x_1,\ldots,x_n) \land \lnot \Phi'(x_1,\ldots,x_n)$.  As the
  size of $\Phi'(x_1,\ldots,x_n)$ is bounded by $n$ plus the sum of
  the sizes of $\varphi''_i$ for $i = 1,\ldots,n$, the formula $\Phi$
  only grows by a linear term in each iteration. As the number of
  iterations is bounded by a polynomial, the maximum size of $\Phi$ is
  also bounded by a polynomial.
  
  It follows that the algoritm is correct, and can be implemented to
  run in polynomial time. \hfill $\Box$
\end{proof}

\begin{corollary} \label{cor:IP}
  There exists a \ptime algorithm that decides whether a Boolean
  combination of linear (in)equalities with rational factors and a
  fixed number of variables has a solution of positive integers.
\end{corollary}
\begin{proof}
  Given a Boolean combination $\Phi(x_1,\ldots,x_n)$ of linear
  (in)equalities with rational factors, we simply apply the algorithm
  of Proposition~\ref{prop:extended-integer-programming} to the
  formula
  $$\Phi'(x_1,\ldots,x_n) = \Phi(x_1,\ldots,x_n) \land \bigwedge_{1 \leq
  i \leq n} x_i \geq 0.$$ \hfill $\Box$
\end{proof}

In the following proposition, we treat the emptiness problem for DTDs:
given a DTD $d$, is $L(d) = \emptyset$? Note that $L(d)$ can be empty
even when $d$ is not. For instance, the trivial grammar $a\to a$
generates no finite trees.  

\begin{proposition} \label{lem:dtd-emptiness-ptime}
  The emptiness problem is (1) \ptime-complete for DTD(NFA) and
  DTD(DFA), and (2) \conp-complete for DTD(SL).
\end{proposition}
\begin{proof}
  (1) The upper bound follows from a reduction to the emptiness
  problem for NTA(NFA)s, which is in \ptime (cf. Theorem 19(1)
  in~\cite{martensneventcs05})
  
  For the lower bound, we reduce from {\sc path
    systems}~\cite{cookpathsys}, which is known to be \ptime-complete.
  {\sc path systems} is the decision problem defined as follows: given
  a finite set of propositions $P$, a set $A \subseteq P$ of axioms, a
  set $R \subseteq P\times P \times P$ of inference rules and some $p
  \in P$, is $p$ \emph{provable from $A$ using $R$}? Here, \emph{(i)}
  every proposition in $A$ is provable from $A$ using $R$ and,
  \emph{(ii)} if $(p_1,p_2,p_3) \in R$ and if $p_1$ and $p_2$ are
  provable from $A$ using $R$, then $p_3$ is also provable from $A$
  using $R$.
  
  In our reduction, we construct a DTD $(d,p)$ such that $(d,p)$ is
  not empty if and only if $p$ is provable.  Concretely, for every
  $(a,b,c)\in R$, we add the string $ab$ to $d(c)$; for every $a\in
  A$, $d(a) = \{\varepsilon\}$.  Clearly, $(d,p)$ satisfies the
  requirements.

  (2) We provide an \np algorithm to check whether a DTD(SL) $(d,r)$
  defines a non-empty language. Intuitively, the algorithm computes
  the set $S = \{a \in \Sigma \mid L((d,a)) \neq \emptyset\}$ in an
  iterative manner and accepts when $r \in S$.

  Let $k$ be the largest integer occurring in any SL-formula in $d$.
  Initially, $S$ is empty. 
  
  The iterative step is as follows. Guess a sequence of different
  symbols $b_1, \ldots, b_m$ in $S$.  Then guess a vector $(v_1,
  \ldots, v_m) \in \{0,\ldots,k+1\}^m$, where $k$ is the largest
  integer occurring in any SL-formula in $d$.  Intuitively, the vector
  $(v_1, \ldots, v_m)$ represents the string $b_1^{v_1} \cdots
  b_m^{v_m}$. From Lemma~\ref{lem:SL-equiv} it follows that any
  SL-formula in $d$ is satisfiable if and only if it is satisfiable by
  a string of the form $a_1^{u_1} \cdots a_n^{u_n}$, where $\Sigma =
  \{a_1, \ldots, a_n\}$, and for all $i = 1, \ldots, n$, $u_i \in
  \{0,\ldots,k+1\}$.  Now add to $S$ each $a \in \Sigma$ for which
  $b_1^{v_1} \cdots b_m^{v_m} \models d(a)$.  Note that this condition
  can be checked in \ptime.  Repeat the iterative step at most
  $|\Sigma|$ times and accept when $r \in S$.

  The \conp-lowerbound follows from an easy reduction of
  non-satisfiability. Let $\phi$ be a propositional formula with
  variables $x_1, \ldots, x_n$. Let $\Sigma$ be the set $\{a_1,
  \ldots, a_n\}$.  Let $(d,r)$ be the DTD where $d(r) = \phi'$, where
  $\phi'$ is the formula $\phi$ in which every $x_i$ is replaced by
  $a_i^{=1}$. Hence, $(d,r)$ defines the empty tree language if and
  only if $\phi$ is unsatisfiable. \hfill $\Box$
\end{proof}

Reducing a grammar is the act of finding an equivalent reduced
grammar.

\begin{corollary} \label{cor:reducing}
  Reducing a DTD(NFA) is \ptime-complete; and reducing a DTD(SL) is
  \np-complete.
\end{corollary}
\begin{proof}
  We first show the upper bounds. Let $(d,s)$ be a DTD(NFA) or DTD(SL)
  over alphabet $\Sigma$. In both cases, the algorithm performs the
  following steps for each $a \in \Sigma$:
  \begin{enumerate}[(i)]
  \item Test whether $a$ is \emph{reachable} from $s$. That is, test whether
    there is a sequence of $\Sigma$-symbols $a_1,\ldots,a_n$ such that
    \begin{itemize}
    \item $a = s$ and $a_n = a$; and
    \item for every $i = 2,\ldots,n$, there exists a string
      $w_1a_{i}w_2 \in d(a_{i-1})$, for $w_1, w_2 \in \Sigma^*$.
    \end{itemize}
  \item Test whether $L((d,a)) \neq \emptyset$.
  \end{enumerate}
  Symbols that do not pass test (i) and (ii) are deleted from the
  alphabet of the DTD. Let $c$ be such a deleted symbol. In the case
  of SL, every atom $c^{\geq i}$ and $c^{=i}$ is replaced by true when
  $i=0$ and false otherwise.  Further, in the case of NFAs, every
  transition mentioning $c$ is removed.

  In the case of a DTD(NFA), step (i) is in \nlogspace and step (ii)
  is in \ptime.  In the case of a DTD(SL), both tests (i) and (ii) are
  in \np.

  For the lower bound, we argue that
  \begin{enumerate}[(1)] 
  \item if there exists an \nlogspace-algorithm for reducing a
    DTD(NFA), then emptiness of a DTD(NFA) is in \nlogspace; and,
  \item if there exists a \ptime-algorithm for reducing a DTD(SL),
    then emptiness of a DTD(SL) is in \ptime.
  \end{enumerate}
  
  Statements (1) and (2) are easy to show: one only has to observe
  that an emptiness test of a DTD can be obtained by reducing the DTD
  and verifying whether the alphabet of the DTD still contains the
  start symbol.  \hfill $\Box$
\end{proof}

\bibliographystyle{plain}
\bibliography{database}

\end{document}